\documentclass[
 reprint,
 amsmath,amssymb,
 aps,nofootinbib,preprintnumbers,longbibliography,superscriptaddress
]{revtex4-2}

\usepackage[utf8]{inputenc}
\usepackage{graphicx}
\usepackage{dcolumn}
\usepackage{bm}
\usepackage{hyperref}
\usepackage{lipsum}
\usepackage{xcolor}
\usepackage{calc}
\usepackage{accents}
\usepackage{comment}
\usepackage{float}
\usepackage{diagbox}
\usepackage{soul}

\makeatletter
\newcommand\footnoteref[1]{\protected@xdef\@thefnmark{\ref{#1}}\@footnotemark}
\makeatother

\newcommand{\LCDM}{$\Lambda$CDM}

\newcommand{\Planck}{{\it Planck}}

\newcommand{\shoes}{S$H_0$ES}

\newcommand{\apjl}{ApJ Letters}
\newcommand{\apjs}{ApJ Supplemental}
\newcommand{\araa}{Ann.~Rev.~Astron.~\& Astrophys.}
\newcommand\mathplus{+}

\begin{document}

\title{Assessing the robustness of sound horizon-free determinations of the Hubble constant}
\author{Tristan L.~Smith}
\affiliation{Department of Physics and Astronomy, Swarthmore College, Swarthmore, PA 19081, USA}
\author{Vivian Poulin}
\affiliation{Laboratoire Univers \& Particules de Montpellier (LUPM), CNRS \& Universit\'e de Montpellier (UMR-5299), Place Eug\`ene Bataillon, F-34095 Montpellier Cedex 05, France}
\author{Th\'eo Simon}
\affiliation{Laboratoire Univers \& Particules de Montpellier (LUPM), CNRS \& Universit\'e de Montpellier (UMR-5299), Place Eug\`ene Bataillon, F-34095 Montpellier Cedex 05, France}

\begin{abstract}
The Hubble tension can be addressed by modifying the sound horizon ($r_{s}$) before recombination, triggering interest in early-universe estimates of the Hubble constant, $H_0$, independent of $r_s$. Constraints on $H_0$ from an $r_s$-free analysis of the full shape BOSS galaxy power spectra within \LCDM{} were recently reported and used to comment on the viability of physics beyond \LCDM. Here we demonstrate that $r_s$-free analyses with current data depend on both the model and the priors placed on the cosmological parameters, such that \LCDM\ analyses cannot be used as evidence for or against new physics. We find that beyond-\LCDM\ models which introduce additional energy density with significant pressure support, such as early dark energy (EDE) or additional neutrino energy density ($\Delta N_{\rm eff}$), lead to larger values of $H_0$. On the other hand, models which only affect the time of recombination, such as a varying electron mass ($\Delta m_e$), produce $H_0$ constraints similar to \LCDM. Using BOSS data, constraints from light element abundances, cosmic microwave background (CMB) lensing, a CMB-based prior on the primordial scalar amplitude ($A_s$), spectral index ($n_s$), and $\Omega_m$ from the Pantheon+ Type Ia supernovae data set, we find that in \LCDM, $H_0=64.9\pm 2.2$ km/s/Mpc; in EDE, $H_0=68.7^{+3}_{-3.9}$; in $\Delta N_{\rm eff}$, $H_0=68.1^{+2.7}_{-3.8}$; and in $\Delta m_e$, $H_0=64.7^{+1.9}_{-2.3}$.  Using a prior on $\Omega_m$ from uncalibrated BAO and CMB  measurements of the projected sound horizon, these values become in \LCDM, $H_0=68.8^{+1.8}_{-2.1}$; in EDE, $H_0=73.7^{+3.2}_{-3.9}$; in $\Delta N_{\rm eff}$, $H_0=72.6^{+2.8}_{-3.7}$; and in $\Delta m_e$, $H_0=68.8\pm 1.9$. With current data, none of the models are in significant tension with \shoes, and consistency tests based on comparing $H_0$ posteriors with and without $r_s$ marginalization are inconclusive with respect to the viability of beyond \LCDM\ models.
\end{abstract}

\maketitle

\section{\label{sec:Intro}Introduction}

As cosmological measurements have become more precise they have revealed a few potential issues within the core cosmological model. This model, referred to as `\LCDM', consists of a geometrically flat universe filled with baryons, photons, three flavors of neutrinos with the standard weak interactions, cold dark matter (CDM), and a cosmological constant, $\Lambda$, with dynamics described by general relativity. The success of this model to describe an exceedingly wide variety of measurements-- from light element abundances produced during big bang nucleosynthesis (BBN), to the cosmic microwave background (CMB), to the clustering of galaxies and the more recent expansion history-- is remarkable (e.g., Refs.~\cite{2018FoPh...48.1226E,1804633}). 

\begin{figure*}[t!] 
   \centering
   \includegraphics[width=0.8\textwidth]{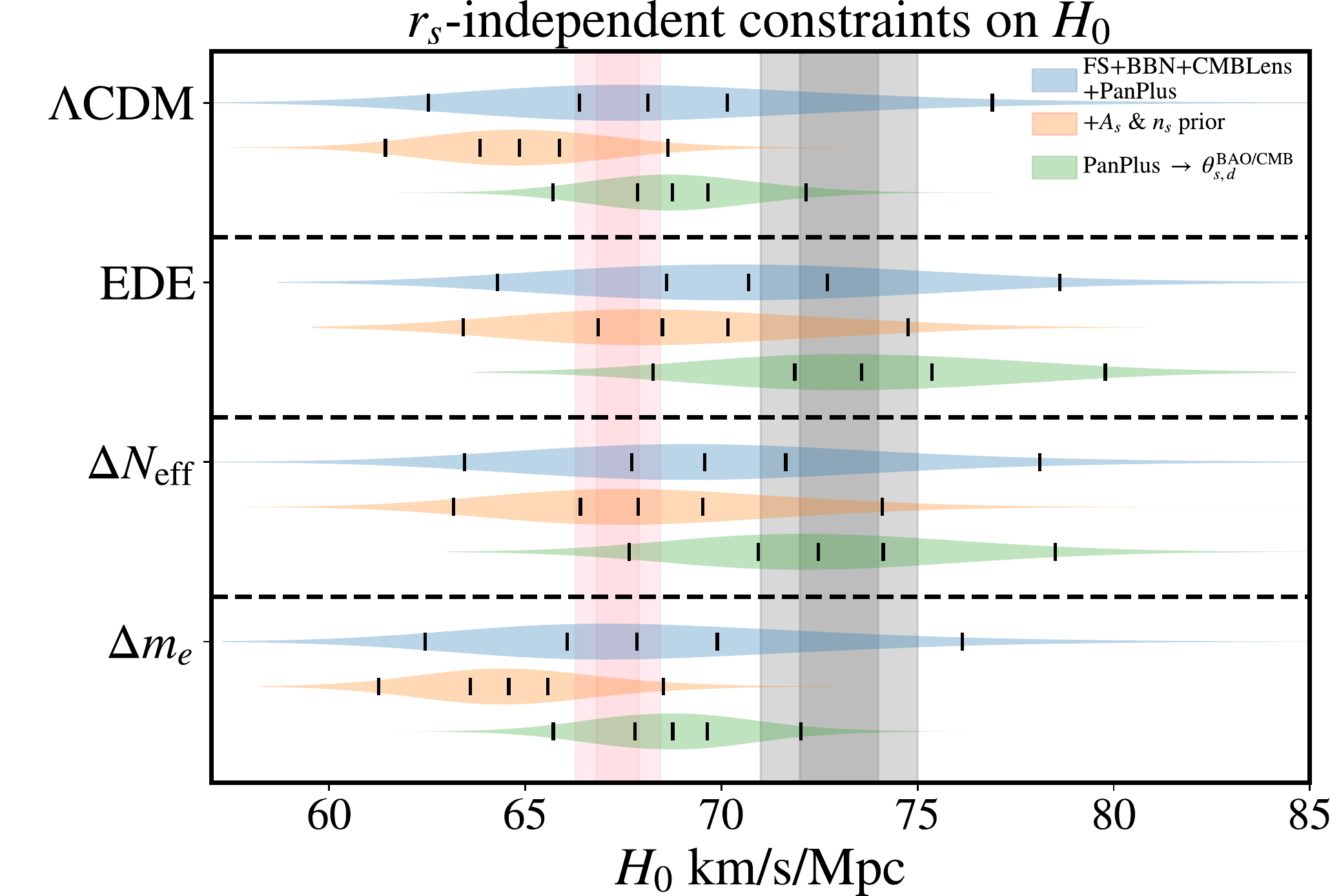} 
   \caption{The 1D posterior distribution for $H_0$ in the four cosmological models we explore here. The central mark shows the mean of the distributions and the outer marks shows the 68\% and 95\% confidence level (CL) regions. The gray band indicates the \shoes~ constraint, $H_0 = 73.04 \pm 1.04$ km/s/Mpc \cite{2022ApJ...934L...7R} and the pink bands the \textit{Planck} value of $H_0=67.36 \pm 0.54$ km/s/Mpc \cite{Planck:2018vyg}. The blue distribution shows the result of an analysis that includes the $r_{s,d}$-marginalized full-shape of the BOSS DR12 galaxy power spectrum (FS), a BBN prior on the baryon density, the \textit{Planck} CMB lensing potential power spectrum (CMBLens), and a prior on $\Omega_m$ from the Pantheon+ Type Ia supernovae data set (PanPlus). The orange distribution shows how constraints shift to lower values when we include a CMB-inspired prior on the scalar amplitude ($A_s$) and slope ($n_s$) and the green curves show how they shift back towards larger values when we additionally replace the $\Omega_m$ prior from Pantheon+ with one from uncalibrated BAO and CMB measurements of the projected sound horizon ($\theta_{s,d}^{\rm BAO/CMB}$). In all cases the distributions for EDE and $\Delta N_{\rm eff}$ are shifted to larger values than in \LCDM\ and $\Delta m_e$.}
   \label{fig:intro_whisker}
\end{figure*}

These measurements allow for a number of non-trivial consistency tests, the most discerning of which allow us to take observations of the early universe (roughly at or before recombination) and \emph{predict} the values of quantities that are measured in the late universe, within a given model. Interestingly, within \LCDM\ applying this to the expansion rate of the universe today, known as the Hubble constant ($H_0$) and to the current amplitude of the clustering of galaxies, quantified by the standard deviation of the mass contained within spheres with radii equal to $8 h^{-1}\ {\rm Mpc}$ ($\sigma_8$), leads to mismatches between the predicted and directly measured values (known as the `Hubble tension' and `$\sigma_8$ tension', respectfully). 
Barring the presence of systematic errors affecting multiple, independent measurements (see Refs.~\cite{Freedman:2021ahq,Riess:2021jrx,Abdalla:2022yfr,Amon:2022azi,Amon:2022ycy} for discussion), this would indicate that one needs to modify \LCDM, and in the process identify new physics that dictate some aspects of the structure and evolution of the universe \cite{Knox:2019rjx,DiValentino:2021izs,Schoneberg:2021qvd}. 

The statistical significance of these mismatches depends on the particular measurements. However, in all cases the value of $H_0$ predicted within \LCDM\ from measurements of pre-recombination physics (the CMB or the baryon acoustic oscillations-- BAO) are smaller than the direct measurements, and in all cases the predicted value of $\sigma_8$ is larger (see, e.g., Refs.~\cite{Freedman:2021ahq,Riess:2021jrx,Riess:2022mme,Abdalla:2022yfr,DiValentino:2021izs,HSC:2018mrq,Heymans:2020gsg,DES:2021wwk}). For individual experiments the mismatch for $H_0$ reaches $\sim 5 \sigma$ (between \textit{Planck} and \shoes~ \cite{Riess:2021jrx,Riess:2022mme}), whereas for $\sigma_8$ it is $\sim 3 \sigma$ (between \textit{Planck} and KiDS-1000 \cite{Heymans:2020gsg}). Regardless of whether or not these discrepancies are due to physics beyond \LCDM\ or yet undiscovered experimental complexities, the increased precision of current cosmological data sets gives us clear motivation to identify additional ways to assess the consistency of \LCDM. 

A fundamentally different type of consistency test focuses on whether a given set of measurements are internally consistent. In the context of CMB measurements, one such approach is to split the data up in multipoles (for the {\it Planck} satellite the split has been typically taken at $\ell \sim 700-800$) and compare the inferred values of the \LCDM\ cosmological parameters \cite{Addison:2015wyg,Planck:2016tof}. Another approach proposes a set of parameters that divides the CMB data into pre- and post-recombination physics \cite{Vonlanthen:2010cd,Audren:2012wb,Audren:2013nwa,Verde:2016wmz}. 

Here we focus on tests based on obtaining constraints on $H_0$ using observations of pre-recombination physics with and without information on the sound horizon, $r_{s}$.\footnote{The sound horizon is time-dependent and hence there are two different values of the sound horizon that impact cosmological measurements: the sound horizon at recombination, $r_{s,{\rm rec}}$, and at baryon decoupling, $r_{s,d}$. The first value is relevant for the CMB and the second for BAO. While the value of either sound horizon can be different in different cosmological models, the difference between them is relatively model-independent with $(r_{s,d}-r_{s,{\rm rec}})H_0 \simeq 6 \times 10^{-4}$ \cite{Lin:2021sfs}.}  In general, determinations of $H_0$ rely on a calibrator, usually in the form of a standard ruler (for CMB/BAO) or a standard candle (for Type Ia supernovae, SNeIa), that breaks the degeneracy between the observed angular size/relative flux of an object, and its true distance to us. In fact, the Hubble tension is often described as a tension between calibrators of the distance ladder, which rely either on the Cepheid variable calibration for the absolute magnitude of SNeIa or the \LCDM\ value of the sound horizon inferred from CMB data \cite{Bernal:2016gxb,Aylor:2018drw}. Consequently, all currently successful attempts to construct `beyond'-\LCDM\ models to address the Hubble tension propose new physics that changes $r_{s}$\footnote{We note that it is not possible to address the Hubble tension by modifying the late-time expansion history \cite{Benevento:2020fev,Efstathiou:2021ocp}.} \cite{Knox:2019rjx,Schoneberg:2021qvd}.
A determination of $H_0$ using observations sensitive to pre-recombination physics which is independent of $r_{s,d}$ (i.e., `$r_{s}$-free') has the potential to provide useful evidence for or against these models \cite{Farren:2021grl}. 

A program of conducting $r_{s}$-free analyses using CMB lensing (along with priors on some of the cosmological parameters) has the potential to achieve this goal, but is fundamentally limited by cosmic variance \cite{Baxter:2020qlr}. It is also possible to use measurements of galaxy clustering, along with an effective marginalization over the value of $r_{s,d}$ \cite{Philcox:2020xbv,Farren:2021grl,Philcox:2022sgj}. Since galaxy surveys have access to a large number of independent modes this has the potential to significantly increase the precision of such an analysis.
To do so, one uses the effective field theory (EFT) of large scale structure \cite{Baumann:2010tm,Carrasco:2012cv,Senatore:2014via,Senatore:2014eva,Senatore:2014vja,Perko:2016puo} applied to the BOSS DR12 galaxy clustering data (EFT BOSS) \cite{BOSS:2016wmc}. The EFT BOSS data have been shown to allow for determination of the $\Lambda$CDM parameters at a precision higher than that from conventional BAO and redshift space distortions, as well as to provide interesting constraints on models beyond $\Lambda$CDM (see, e.g., Refs.~\cite{DAmico:2019fhj,Ivanov:2019pdj,Colas:2019ret,DAmico:2020kxu,DAmico:2020tty,Chen:2021wdi,Zhang:2021yna,Zhang:2021uyp,Philcox:2021kcw,Simon:2022ftd,Kumar:2022vee,Nunes:2022bhn,Lague:2021frh,Carrilho:2022mon,Simon:2022adh}).

The way in which $r_{s}$-free inferences of $H_0$ may impact models that attempt to resolve the Hubble tension is two-fold. First, as a predictive test, it could indicate that models which alter $r_{s}$ to address the Hubble tension are disfavored if the $r_{s}$-independent value of $H_0$ is in tension with direct measurements of $H_0$ \cite{Philcox:2022sgj}. Second, as an internal consistency test, a comparison between constraints to $H_0$ with and without $r_{s,d}$ can serve as an indicator for or against beyond-\LCDM\ physics \cite{Farren:2021grl}. 

Here we explore whether these analyses provide a robust test of new physics by considering three \LCDM\ extensions which affect $r_{s,d}$: an axion-like model of early dark energy (EDE), a model with additional free-streaming ultra-relativistic energy density ($\Delta N_{\rm eff}$), and a model with a value of the electron mass which is different at recombination than it is today ($\Delta m_e$). We also investigate how various external priors affect these results. Fig.~\ref{fig:intro_whisker} summarizes the 1D posteriors of $H_0$ in the $r_{s,d}$-marginalized analysis for the four models considered in this work.

Using an $r_{s}$-free analysis of BOSS DR12, \textit{Planck} CMB lensing, a BBN prior, and $\Omega_m$ estimated from Pantheon+ \cite{Brout:2022vxf}, we find that both $\Delta N_{\rm eff}$ and EDE open up a new degeneracy between $H_0$ and the primordial power spectrum (i.e., the scalar amplitude, $A_s$, and index, $n_s$) leading to a posterior distribution for $H_0$ that is shifted to higher values compared to \LCDM.  We find that for all four models we consider, the posterior for $H_0$ is consistent with the \shoes~ determination of $H_0$ at $\sim 1.5\sigma$\footnote{In this paper we quote tension assuming Gaussian posteriors for simplicity. This slightly overestimates the level of tension, due to long tails of distribution, but does not affect our conclusions.}. When imposing an additional CMB-inspired prior on the primordial power spectrum, the inferred value of $H_0$ in \LCDM\ and $\Delta m_e$ is in tension with \shoes~ at $\sim 3.5\sigma$, whereas for $\Delta N_{\rm eff}$ and EDE the tension drops to $1.7 \sigma$ and $1.3 \sigma$, respectively. As a result we find that, the value of $H_0$ infered from an $r_{s,d}$-marginalized analysis is model dependent. We also find that, as an internal consistency test, with and without $r_{s,d}$ marginalization, the $H_0$ posteriors are in statistical agreement for all of the models we consider. 

This paper is organized as follows. In Sec.~\ref{sec:scalings} we establish the way in which the various quantities in the $r_{s}$-free data depend on \LCDM\ parameters. This allows us to anticipate the various degeneracies in a full analysis of these data and establish the role played by $A_s$ and $n_s$ in constraining $h$. In Sec.~\ref{sec:data} we describe the data sets we use as well as the Markov Chain Monte Carlo (MCMC) analysis we perform. In Sec.~\ref{sec:LCDM} we establish that within \LCDM, constraints to $h$ are driven by measurements of the amplitude of the matter power spectrum and constraints to $\Omega_m h^p$ with $1\lesssim p \lesssim 2$. In Sec.~\ref{sec:beyondLCDM} we perform $r_{s}$-free analyses on three beyond-\LCDM\ models. We conclude and discuss the implications of our results in Sec.~\ref{sec:conclusions}. In Appendix \ref{app:scaling} we give details about how the various data we use depends on cosmological parameters, in Appendix \ref{app:peak} we demonstrate that the peak of the matter power spectrum does not play a significant role in constraining the Hubble constant, in Appendix \ref{app:check} we demonstrate that the broadband/BAO split algorithm works even in cases where the sound horizon deviates significantly from the \LCDM\ value, and in Appendix \ref{app:PanPlus_check} we show that for the models we consider the full Pantheon+ likelihood is well captured by using a prior on $\Omega_m$. 

\section{$H_0$ from galaxy clustering and CMB lensing}
\label{sec:scalings}

To build an intuition as to how $h$ can be constrained without the sound horizon, it is helpful to establish the approximate relationship between the galaxy power spectrum/CMB lensing and the \LCDM\ parameters whose values we infer from these data. 
In this discussion we make the important distinction between how the \emph{amplitude} (i.e., $k$-independent part) and \emph{shape} (i.e., $k$-dependent part) of the galaxy power spectrum provides information about the Hubble constant. The work in Refs.~\cite{Philcox:2020xbv,Farren:2021grl, Philcox:2022sgj} emphasizes the role that the shape of the galaxy power spectrum plays-- in particular the wavenumber which enters the horizon at matter/radiation equality $k_{\rm eq}$. Here we show that the amplitude of the $k>k_{\rm eq}$ part of the galaxy power spectrum also plays an important role in constraining $h$. 

The basic shape of the galaxy power spectrum is set by two main scales: the sound horizon at baryon decoupling
\begin{equation}
    r_{s,d} \equiv \int_{z_d}^{\infty} \frac{c_s(z')}{H(z')} dz', \label{eq:rs}
\end{equation}
where $z_d$ is the redshift at which baryons decouple and $c_s(z)$ is the photon/baryon sound speed (see, e.g., Ref.~\cite{Planck:2013pxb})
and the wavenumber which enters the horizon at matter/radiation equality,
\begin{equation}
    k_{\rm eq} = \frac{\omega_m}{h\sqrt{\omega_r/2}} \frac{100\ h {\rm km/s/Mpc}}{c},
\end{equation} 
where the last term comes from introducing $h \equiv H_0/(100 \ {\rm km/s/Mpc})$.

The effects of baryons are imprinted through $r_{s,d}$ and an additional scale, $k_d \equiv H(z_d)/(1+z_d)$, the size of the horizon when baryons decouple from photons and start to fall into the gravitational potentials. The largest effect is a suppression of power at wavenumbers larger than $k_d$ compared to a CDM-only universe. The acoustic oscillations in the baryon/photon fluid (i.e., the BAO) are also imprinted into the galaxy power spectrum as oscillations with a frequency set by integer multiples of $k_{s,d} \equiv 2\pi/r_{s,d}$ \cite{Brieden:2021edu}.  We note that since $k_d < k_{\rm eq} \simeq 0.01\ h{\rm Mpc^{-1}}$ this scale is too large to be probed with current galaxy surveys.

The value of $k_{\rm eq}$ plays two important roles in the galaxy power spectrum: it sets the wavenumber at the peak, as well as the range of scales experiencing a logarithmic enhancement in power at $k>k_{\rm eq}$. In practice, measurements of the galaxy power spectrum cannot probe scales large enough to get a precise measure of the location of the peak \cite{Philcox:2020xbv} (though we note that future HI surveys will be able to measure the peak \cite{Cunnington:2022ryj}). For the main analysis presented here we take $k_{\rm min} = 0.01\ h{\rm Mpc}^{-1}$ which is just slightly smaller than the typical values of $k_{\rm eq}$. In Appendix \ref{app:peak} we also perform an analysis with a larger $k_{\rm min}$ in order to demonstrate that the location of the peak of the galaxy power spectrum does not play a dominant role in constraining $h$. Because of this, most of the sensitivity to $k_{\rm eq}$ is not in the peak of the galaxy power spectrum, but from the amplitude at scales $k>k_{\rm eq}$ \cite{Philcox:2020xbv}.

Yet, the measurements of $h$ do not only rely on $r_{s,d}$ and $k_{\rm eq}$, but also on the overall amplitude of the galaxy power spectrum. As discussed in more detail in Appendix \ref{app:scaling}, the galaxy power spectrum amplitude reflects the fact that during radiation domination Hubble friction limits the growth of dark matter perturbations. Once radiation domination ends, the dark matter perturbations grow proportional to the scale factor, $a$. Therefore, the amplitude of the matter power spectrum scales with $(a/a_{\rm eq})^2 \propto a^2 \Omega_m^2 h^4$. In this way information about $h$ contained in the amplitude of the galaxy power spectrum provides us with a `standard clock', measuring how much the dark matter perturbations have grown since matter/radiation domination. 

Summarizing the results of Appendix \ref{app:scaling}, we can write how the galaxy power spectra and CMB lensing potential power spectrum depend on the amplitude of the primordial power spectrum, $A_s$, the normalized Hubble constant, $h$, the `geometric' matter density, $\Omega_m$, and the physical radiation energy density, $\omega_r \equiv \Omega_r h^2$. Note that all measured quantities are dimensionless, so in the following equations lengths are written in  $h^{-1}\ {\rm Mpc}$. 

The overall amplitudes of the galaxy power spectrum, $P_{\rm gal}$, and the CMB lensing power spectrum, $C_L^{\phi \phi}$, scale as 
\begin{eqnarray}
P_{\rm gal} &\propto& b^2 R_c^2 A_s\Omega_m^{2.25} \omega_r^{-2}\boldsymbol{h^4}  , \label{eq:galAmp}\\
L^4 C_L^{\phi \phi} &\propto& A_s  \Omega_m^{3.5} \omega_r^{-1}\boldsymbol{h^{2.6}}, \label{eq:lensAmp}
\end{eqnarray}
where $b$ is the linear bias, $R_c \equiv \omega_{cdm}/\omega_m = 1-\omega_b/\omega_m$ is the baryon suppression \cite{Bernal:2020vbb}, and $\omega_{cdm}$ is the physical cold dark matter density today. 
It is interesting to note that, while these are all proportional to $A_s$, they depend on different powers of  with $h$ indicating that the combination of $P_{\rm gal}$ and $C_L^{\phi \phi}$ can break the $A_s-h$ degeneracy. It is also evident that additional information on $\Omega_m$ will further help constraining $h$.

The shape of these power spectra depend on
\begin{eqnarray}
\left(\frac{k}{k_p/h}\right)^{n_s-1}&=&\left(\frac{k}{(0.05/{\bf h})\ h{\rm  Mpc}^{-1}}\right)^{n_s-1},\\
k_{\rm eq}/h &\propto& \Omega_m\omega_r^{-0.5}\boldsymbol{h},\\
\ell^{\phi \phi}_{\rm peak} &\propto& \Omega_m^{0.75}\omega_r^{-0.5}\boldsymbol{h}\,,
\end{eqnarray}
where $n_s$ is the primordial scalar spectral index and $k_p = 0.05\ {\rm Mpc}^{-1}$ is the standard pivot scale \cite{Planck:2013pxb}. Note that the different $\Omega_m$ scalings provide a way to break the degeneracy between $\Omega_m$ and $h$. Moreover, the baryon suppression and amplitude of the BAO in the galaxy power spectrum gives information about the ratio $\omega_b/\omega_{m}$. 
Finally, redshift space distortions provide additional sensitivity to
\begin{equation}
f \sigma_8 \propto A_s^{1/2}  \Omega_m^{1.25} \omega_r^{-0.65} \boldsymbol{h^{1.75}},\label{eq:rsd}
\end{equation}
where $f$ is the growth rate and $\sigma^2_8$ is the variance of the fractional mass fluctuations in spheres of comoving radius $R = 8 h^{-1}\ {\rm Mpc}$. 

These scaling equations allow us to understand general trends in the posterior distributions. First, as it was noted before, it is clear that knowledge of $k_{\rm eq}\propto \Omega_m h$ and $\Omega_m$ from, e.g. SNeIa, provides a constraint on $h$ \cite{Philcox:2020xbv}. In fact, the whole shape of the power spectra, through $k_{\rm eq}$, baryon suppression,  and $\ell_{\rm peak}^{\phi \phi}$, provides constraints to $\Omega_m h^p$, where $1 \lesssim p \lesssim 2$.  Yet, these are not the only parameters appearing in this scaling: $A_s$ and $n_s$ also play an important role. For example, in \LCDM\, $\omega_r$ is fixed and with a Pantheon+ prior on $\Omega_m$, Eqs.~(\ref{eq:galAmp}), (\ref{eq:lensAmp}) and  (\ref{eq:rsd}) show that an increase in $h$ must be accompanied by a decrease in $A_s$ in order to keep the amplitudes unaffected. This additional degeneracy will be even more important for beyond $\Lambda$CDM determinations of $h$.

\section{Cosmological models and data analysis}
\label{sec:data}

In order to explore the extent to which an $r_{s}$-free analysis may depend on the cosmological model, we consider three beyond-\LCDM\ models that affect the value of the sound horizon in different ways. 

The sound horizon is inversely proportional to the Hubble parameter before recombination [see Eq.~(\ref{eq:rs})]. We consider two models which lead to changes in the early-universe Hubble parameter: variations in the number of ultra-relativistic neutrinos, $\Delta N_{\rm eff}$ (we always take one neutrino to have a mass of 0.06 eV), and the ultra-light axion-inspired model for EDE \cite{Poulin:2018cxd,Smith:2019ihp} (we use the scalar field potential $V = m^2 f^2 [1-\cos(\phi/f)]^3$, where $m$ is the axion mass, $f$ is the axion decay constant and $\phi$ the field value). As described in Ref.~\cite{Smith:2019ihp}, we use a shooting method to map the set of phenomenological parameters $\{\log_{10}(z_c), f_{\rm EDE}(z_c)\}$ (which describe when the field becomes dynamical and its maximum fractional contribution to the total energy density, respectively) to the theory parameters $\{m,f\}$. A major difference between these two `energy density modification' models is that while a change to the neutrino energy density has an impact throughout radiation domination, the EDE's energy density makes a dynamically relevant contribution to the total energy density over a relatively short period of time. 

The sound horizon depends on the redshift at which baryons decouple from photons, $z_d$ [see Eq.~(\ref{eq:rs})]. We also consider a model in which the mass of the electron may be different around recombination than its value today, leading to a change in the Thomson scattering cross section, and hence changing $z_d$ (see, e.g., Refs.~\cite{Hart:2017ndk,Hart:2021kad}). 

Our Markov-Chain Monte Carlo (MCMC) analyses uses \texttt{MontePython-v3}\footnote{\url{https://github.com/brinckmann/montepython_public}} code \cite{Audren:2012wb,Brinckmann:2018cvx} interfaced with modified versions of \texttt{CLASS-PT}\footnote{\url{https://github.com/Michalychforever/CLASS-PT}} which is itself a modified version of \texttt{CLASS}\footnote{\url{https://lesgourg.github.io/class_public/class.html}} \cite{Blas:2011rf}. 

In this paper, we carry out various analyses using a combination of the following data sets:
\begin{itemize}

    \item {\bf Full-shape galaxy power spectra (FS):} The effective field theory (EFT) of large scale structure applied to the BOSS DR12 galaxy clustering data. For the main analysis we use the same data and code as in Ref.~\cite{Philcox:2022sgj}: we use the power spectrum measured in Ref.~\cite{Philcox:2021kcw} from the $z=0.38$ and 0.61 redshift bins at the Northern and Southern Galactic Caps \cite{2015ApJS..219...12A}. We use the unreconstructed monopole, quadrupole, and hexadecapole galaxy power spectrum multipoles with\footnote{For one part of our analysis we increase the minimum $k$ to $0.05\ h{\rm Mpc}^{-1}$ for the galaxy power spectrum multipoles.} $0.01 h\ {\rm Mpc}^{-1} \leqslant k \leqslant 0.2h \ {\rm Mpc}^{-1}$ and the real-space extension, $Q_0$, with $0.2\ h {\rm Mpc}^{-1} \leqslant k \leqslant 0.4\  h {\rm Mpc}^{-1}$. We include EFT parameters and priors as described in Refs.~\cite{Philcox:2021kcw,Simon:2022lde}. Note that these priors were shown to be informative, and part of our results could be affected by the choice of priors, at the $1\sigma$ level \cite{Simon:2022lde}, but we do not expect our main conclusions to change.
    
    \item {\bf BBN:} The BBN measurement of $\omega_b$ \cite{Schoneberg_2019} that uses the theoretical prediction of \cite{Consiglio_2018}, the experimental Deuterium fraction of \cite{Cooke_2018} and the experimental Helium fraction of \cite{Aver_2015}. Note that this likelihood also tightly constrains $\Delta N_{\rm eff}$ \cite{Schoneberg_2019}. As we are interested in computing constraints driven by galaxy clustering/CMB lensing, when varying $\Delta N_{\rm eff}$ we instead use a Gaussian prior on $\omega_b = 0.02268 \pm 0.00038$ \cite{Ivanov:2019pdj}.
    \item {\bf CMB lensing (CMBLens):} The CMB-marginalized gravitational lensing potential from \Planck{} 2018 temperature and polarization data with $8\leqslant L \leqslant 400$ \cite{Planck:2018lbu}. 
    \item {\bf Pantheon+ (PanPlus):} The Pantheon+ measurement of $\Omega_m=0.338 \pm 0.018$ using uncalibrated Type Ia supernovae (SNeIa), modeled as a Gaussian likelihood \cite{Brout:2022vxf}. We have explicitly checked that this prior captures all of the information contained within the full likelihood in Appendix \ref{app:PanPlus_check}. 
    \item {\bf Uncalibrated BAO and CMB  measurements of the projected sound horizon ($\boldsymbol{\theta_{s,d}^{\rm BAO/CMB}}$):} In some of our analyses we have replaced the $\Omega_m$ prior from PanPlus with one from an analysis of uncalibrated BAO measurements of $r_{s,d}H(z)$ and $\theta_{s,d}(z) =r_{s,d}/D_A(z)$ and the \textit{Planck} inferred value of $100\theta_{s,d}(z_{\rm CMB})$, $\Omega_m = 0.30 \pm 0.01$\cite{Lin:2021sfs}. We have taken the uncertainty to be 25\% larger in order to account for variations in $\theta_{s,d}(z_{\rm CMB})$ when fit to the cosmological models we consider. We note that some of the data used to generate this prior is correlated with the FS data, but stress that our use of this prior is primarily meant to highlight how constraints on $\Omega_m$ affect the $r_s$-free results.
    \item {\bf CMB priors:} For some of our analyses we use the Gaussian priors $\ln 10^{10} A_s = 3.044\pm0.08$ and $n_s = 0.96 \pm 0.03$. The prior on $A_s$ is 8\% around the \textit{Planck} mean value \cite{Baxter:2020qlr,Philcox:2022sgj} and the prior on $n_s$ is based on the one used in Ref.~\cite{Philcox:2022sgj}, but is lightly wider in order to account for the fact that some of the beyond-\LCDM\ models we consider, when fit to the CMB, lead to larger values for $n_s$ (see, e.g, Refs.~\cite{Smith:2019ihp,Ye:2021nej,Aloni:2021eaq}).
\end{itemize}

In the following we denote the combination of FS, BBN, CMBLens, and PanPlus as `All', to distinguish it from analyses that just combine a subset of these data sets. 

All MCMCs use wide uninformative flat priors on the physical CDM energy density, $\omega_{cdm}$, the Hubble parameter today in units of 100km/s/Mpc, $h$, the logarithm of the variance of curvature perturbations centered around the pivot scale $k_p = 0.05$ Mpc$^{-1}$ (according to the \Planck{} convention \cite{Planck:2013pxb}), $\ln 10^{10}A_s$, and the scalar spectral index $n_s$.

We marginalize over information about the sound horizon in the galaxy power spectra following the procedure introduced in Ref.~\cite{Farren:2021grl}. This involves splitting the linear power spectrum into its broadband (BB) shape and the BAO and marginalizing over a new scaling parameter, $\alpha_{r_s}$, 
\begin{equation}
    P_{\rm lin}(k) = P_{\rm BB}(k) + P_{\rm BAO}(\alpha_{r_s} k).
\end{equation}
As with the cosmological parameters, we use a wide uninformative flat prior on $\alpha_{r_s}$. We note that Ref.~\cite{Philcox:2022sgj} places a Gaussian prior with mean equal to 1 and a standard deviation of 0.5. Since the value of $\alpha_{r_s}$ only varies by $\sim 0.1$ their choice of prior is also uninformative. 

For the three free parameters of the EDE model, we impose a logarithmic priors on $z_c$, and flat priors for $f_{\rm EDE}(z_c)$ and $\theta_i$:
\begin{align*}
    &3 \le \log_{10}(z_c) \le 4, \\
    &0 \le f_{\rm EDE}(z_c) \le 0.5, \\
    &0 \le \theta_i\equiv \phi_i/f \le 3.1.
\end{align*}
When we vary the electron mass we use the prior $0.8 \leqslant m_e/m_{e,0} \leqslant 1.2$, while we take $\Delta N_{\rm eff} \geqslant 0$ when we vary the amount of free-streaming ultra-relativistic energy density.
We define our MCMC chains to be converged when the Gelman-Rubin criterion $R-1 < 0.05$ \cite{Gelman:1992zz}. Finally, we produce our figures using \texttt{GetDist} \cite{Lewis:2019xzd}.

\section{Constraints on $h$ in \LCDM}
\label{sec:LCDM}

We start by comparing the 1D posterior distributions of $h$ from analyzing FS+BBN+CMBLens+PanPlus (the `All' dataset), with and without marginalizing over $r_{s,d}$, without applying any CMB priors on $A_s$ or $n_s$.
The \LCDM\ posterior distributions are summarized in Tab.~\ref{tab:LCDM}, and we find:
 \begin{eqnarray}
  h &=&0.697^{+0.014}_{-0.016}\,~{\rm w/o}~r_{s,d} {\rm -marg}.,\nonumber\\
    h &=&0.687^{+0.030}_{-0.050}\,~{\rm w/}~r_{s,d} {\rm -marg}.\nonumber\,
 \end{eqnarray}
which shows no significant tension with \shoes~ even when marginalizing over $r_{s,d}$. We note that, as shown in Table \ref{tab:LCDM} the mean value of $r_{s,d}$ is $\sim 5$ Mpc smaller than the value preferred by \textit{Planck} \cite{Planck:2018vyg}. This is due to the fact that these data prefer a significantly larger mean physical CDM density, $\omega_{cdm} \sim 0.14$, compared to \textit{Planck}, $\omega_{cdm} \sim 0.12$. The larger $\omega_{cdm}$, combined with the relatively large value of $\Omega_m$ from PanPlus, leads to the statistical agreement between constraints to $h$ from the two datasets. We also note that we find no significant shift between the two values of $h$, which has been advocated as being a hint against the presence of new physics affecting the sound horizon \cite{Philcox:2021kcw,Philcox:2022sgj}.
\begin{figure}[h!] 
   \centering
   \includegraphics[width=\columnwidth]{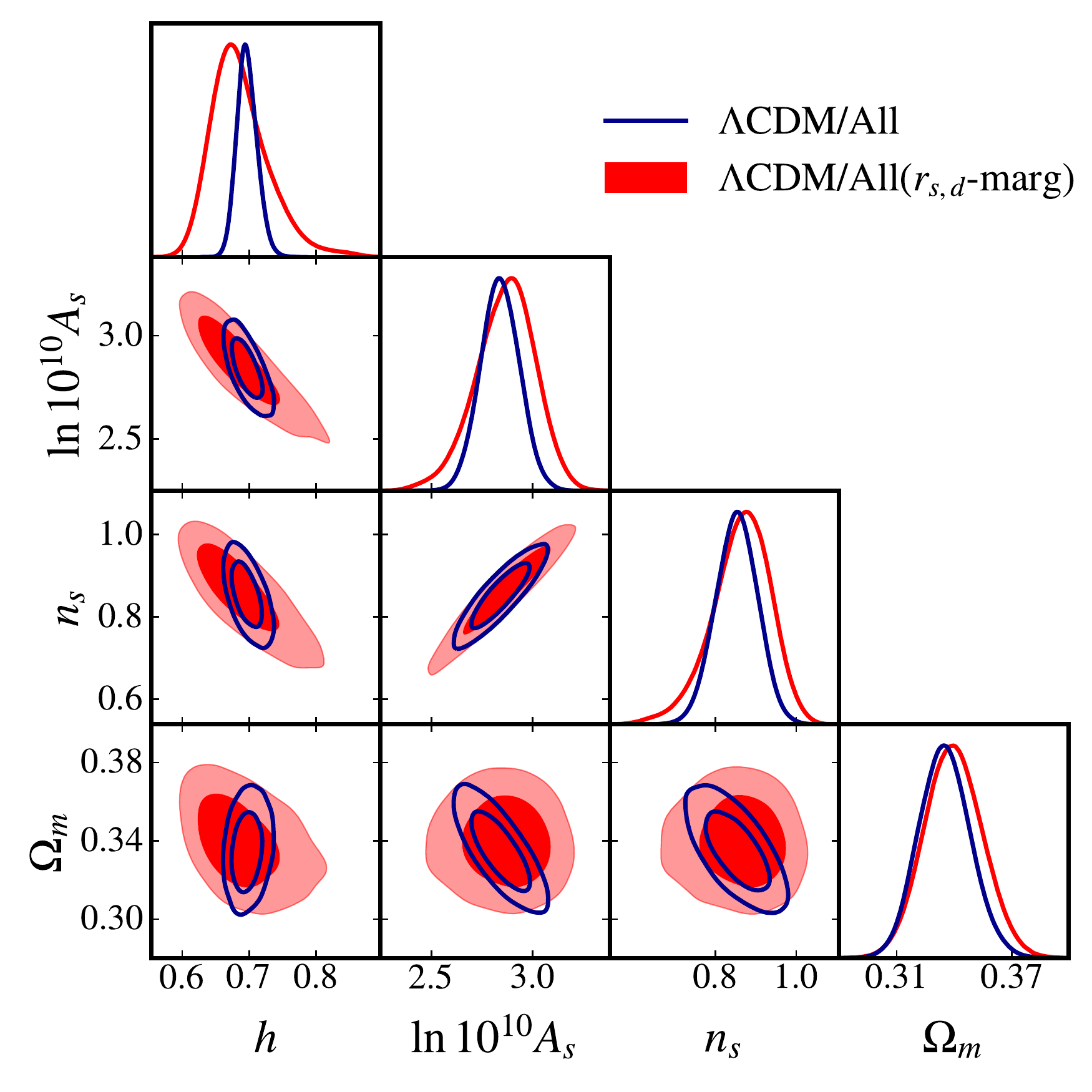} 
   \caption{A triangle plot showing the constraints to four of the five \LCDM\ cosmological parameters with and without marginalization over $r_{s,d}$ (we do not show the posterior distribution for $\omega_b$ since it is well constrained by the BBN likelihood). The filled contours show constraints using `All' of the data without marginalization over $r_{s,d}$ whereas the unfilled contours show the same data with marginalization over $r_{s,d}$.}
   \label{fig:LCDM_comp}
\end{figure}
\begin{table}
\def\arraystretch{1.0}
    \scalebox{1.0}{
    \begin{tabular}{|l|c|c|} 
        \hline
        Parameter & \LCDM\ (no $r_{s}$-marg) & \LCDM\ ($r_{s}$-marg) \\
        \hline
        \hline
        $10^{2}\omega{}_{b }$ & $2.273\pm 0.038$ & $2.273\pm 0.037$ \\
        $\omega{}_{cdm }$&$0.1395^{+0.0091}_{-0.012}$ & $0.137^{+0.011}_{-0.022}$\\
        $\Omega{}_{m }$ &  $0.335\pm 0.013$ & $0.340\pm 0.015$ \\
        $h$  &  $0.697^{+0.014}_{-0.016}$ & $0.687^{+0.030}_{-0.050}$ \\
        $\ln 10^{10}A_s$&  $2.839\pm 0.096$ & $2.86^{+0.16}_{-0.13}$ \\
        $n_{s }$&$ 0.853\pm 0.052$  & $0.863^{+0.081}_{-0.060}$ \\
        $r_{s,d}$ [Mpc] & $141.9^{+2.7}_{-2.4}$  & $142.7^{+5.1}_{-3.2}$ \\
        $\alpha_{r_s}$ &-- & $1.011^{+0.036}_{-0.028}$\\
        \hline
    \end{tabular} }
    \caption{The mean and $\pm 1\sigma$ uncertainties of the \LCDM\ cosmological parameters with and without marginalization over $r_{s,d}$ and using `All' of the data.}
    \label{tab:LCDM}
\end{table}
This $r_{s,d}$-marginalized value is larger than the main value reported in Ref.~\cite{Philcox:2022sgj} because here we have not imposed any external priors on $n_s$ or $A_s$.\footnote{Ref.~\cite{Philcox:2022sgj} argues that their results are robust to dropping any priors on $n_s$ or $A_s$, in this case reporting $h= 0.660^{+0.027}_{-0.034}$. However, even after adopting the same parameter settings as they use, without these priors we find $h=0.677_{-0.037}^{+0.028}$, giving posterior on $h$ that is consistent with the \shoes~ value at $\sim 2 \sigma$.} As shown in Table \ref{tab:LCDM}, both $A_s$ and $n_s$ are lower than what is found using CMB data, and when imposing priors from the CMB the posterior on $h$ can change appreciably. In Sec.~\ref{sec:As/ns} we explore the degeneracy between $h$ and $A_s/n_s$ and in Sec.~\ref{sec:LCDM_priors} we show the impact of imposing the CMB priors.  

\subsection{The $A_s/n_s$-degeneracy}
\label{sec:As/ns}

To understand the role that the FS galaxy power spectra are playing in constraining $h$, Fig.~\ref{fig:LCDM_comp} shows a comparison between two different data analyses: FS+BBN+CMBLens+PanPlus with and without $r_{s}$-marginalization. First, focusing on the constraints to $h$ (left-most column) and on the analysis without $r_{s,d}$-marginalization one can see that the constraint on $h$ is less degenerate with $A_s$, $n_s$, and $\Omega_m$ than with $r_{s,d}$-marginalization. This shows that when including information on $r_{s,d}$ one gains independent information on $h$ through its effect on the projected size of the sound horizon. When marginalizing over $r_{s,d}$ on the other hand, one can see that $h$ is anti-correlated with $A_s$/$\Omega_m$, as expected from the discussion in Sec.~\ref{sec:scalings}. In particular the degeneracy between $h$ and $A_s$ provides evidence that constraints on $h$ when marginalizing over $r_{s,d}$, at least in part, come from the amplitude of the galaxy power spectrum. 

\begin{figure}[t!] 
   \centering
   \includegraphics[width=0.9\columnwidth]{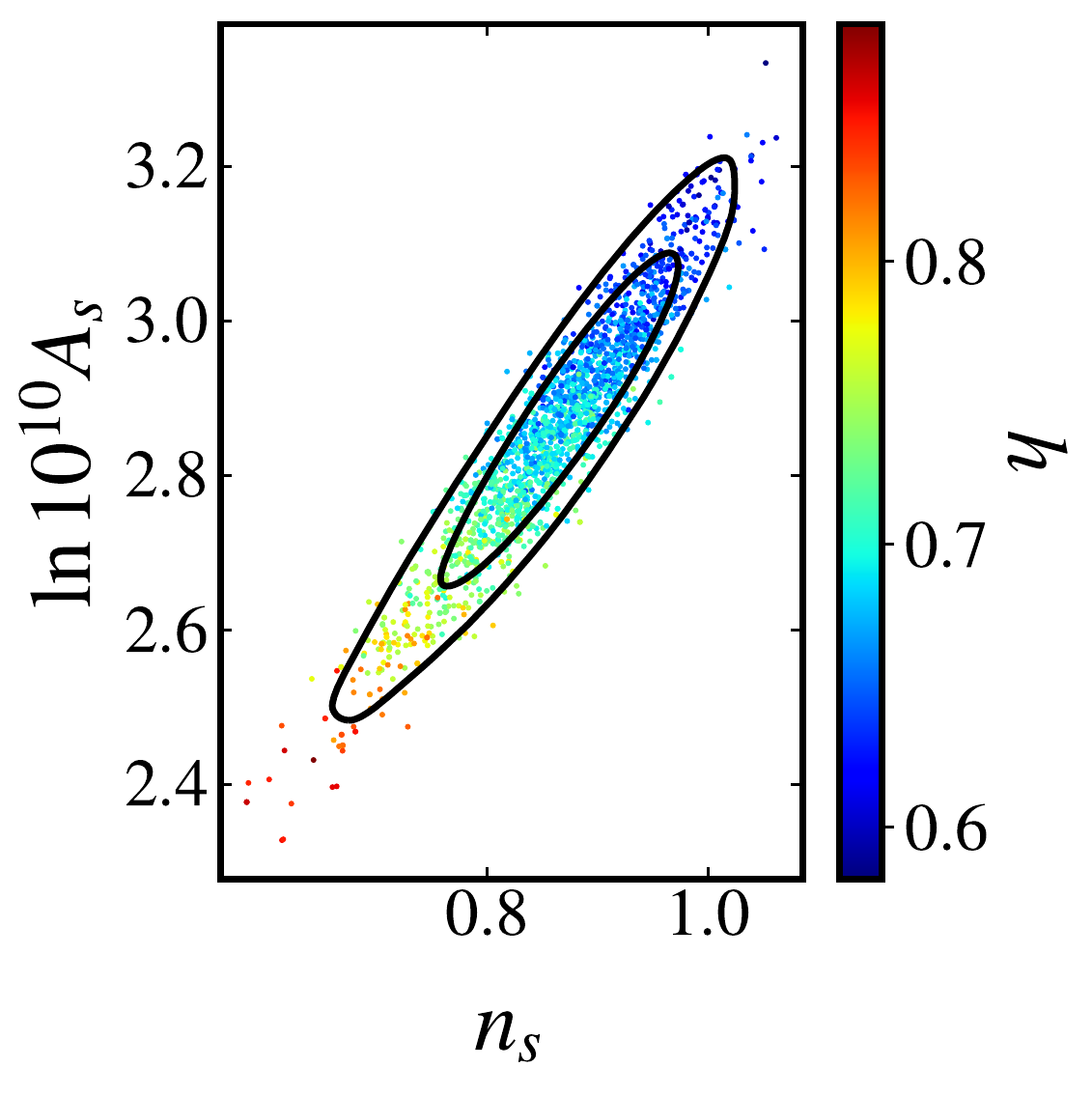} 
   \caption{The 3D correlation between $n_s$, $\ln 10^{10} A_s$, and $h$ for an $r_{s,d}$-marginalized analysis of \LCDM\ using `All' of the data. We can see that $h$ has a strong negative correlation with both $A_s$ and $n_s$.}
   \label{fig:LCDM_ns_vs_h}
\end{figure}

Fig.~\ref{fig:LCDM_comp} clearly shows that when marginalizing over $r_{s,d}$, $h$ and $n_s$ are anti-correlated. This anticorrelation is also related to the primordial amplitude of the fluctuations which can be seen in the 3D plot in Fig.~\ref{fig:LCDM_ns_vs_h}. There we can see that a decrease in $n_s$ is compensated by a decrease in $A_s$ and an increase in $h$. This relationship is due to a balance between the enhancement of power for $k>k_{\rm p} = 0.05/h\ h{\rm Mpc^{-1}}$ (for $n_s<1$) and the shift with $h$ in scale at which the logarithmic enhancement starts, $k_{\rm eq} = \Omega_m h$.  

\begin{figure}[h!] 
  \centering
  \includegraphics[width=\columnwidth]{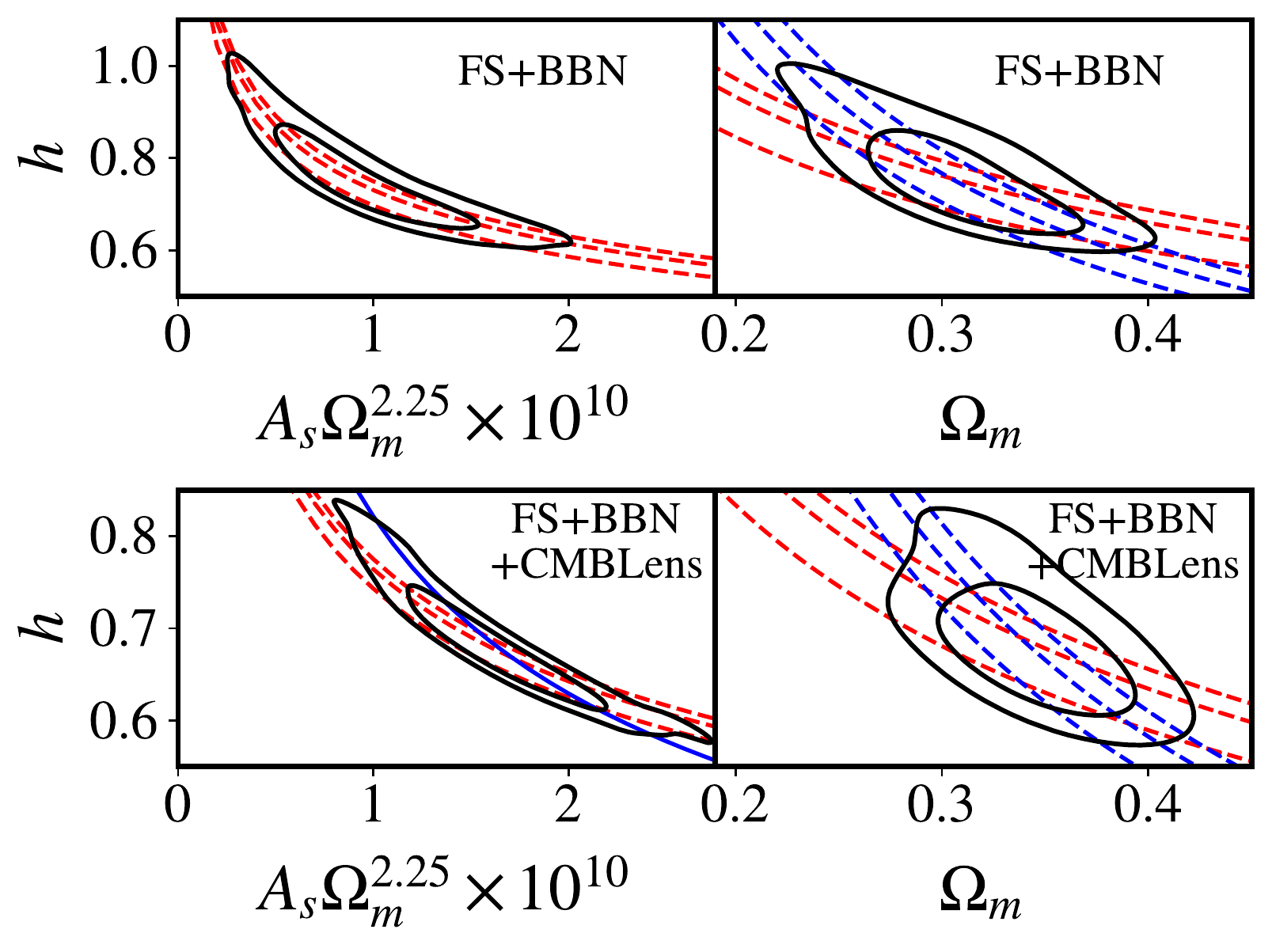} 
  \caption{A pair of 2D posteriors comparing $r_{s}$-marginalized constraints using FS + BBN (top row) vs.~ FS+BBN+CMBLens. The red curves in the two left panels show the mean and $\pm 1 \sigma$ values for $A_s \Omega_m^{2.25} h^4$, which captures information in the small-scale ($k> k_{\rm eq}$) part of the galaxy power spectrum [see Eq.~(\ref{eq:galAmp})]. The solid blue curve in the bottom left panel corresponds to the $\propto h^{2.6}$ scaling from the lensing data [see Eq.~(\ref{eq:lensAmp})]. The red/blue curves in the right panels show the mean and $\pm 1 \sigma$ for $\omega_m = \Omega_m h^2$ and $k_{\rm eq} = \Omega_m h$, respectively. The agreement between the curves and the contours indicates that both the small-scale amplitude and the various combinations of $\Omega_m h^p$ (characterizing the shape of the power spectra), with $1\lesssim p \lesssim 2$, plays a role in constraining $h$.}
  \label{fig:h_degen}
\end{figure}

We further explore the exact shape of the degeneracies introduced through the amplitudes of the $k>k_{\rm eq}$ galaxy power spectra and the CMB lensing potential power spectrum as discussed in Sec.~\ref{sec:scalings} and Appendix \ref{app:scaling}. The top left panel of Fig.~\ref{fig:h_degen} shows the results of using FS + BBN. The dashed red curves show the mean and $\pm 1 \sigma$ of $A_s \Omega_m^{2.25} h^4$, which sets the $k>k_{\rm eq}$ amplitude of the galaxy power spectrum [see Eq.~(\ref{eq:galAmp})]. The agreement between the red curves and the 2D posterior definitively demonstrates the importance of the $k>k_{\rm eq}$ amplitude of the galaxy power spectrum in constraining $h$ with these data. 

The bottom left panel of Fig.~\ref{fig:h_degen} shows the same curve/2D posteriors but with FS+BBN+CMBLens. There we can see that the addition of CMB lensing data shifts the $h$ vs.~$A_s \Omega_m^{2.25}$ contour, and decreases the width of the posterior, indicating that CMB lensing adds information on $h$. The blue curve in this panel shows the $\propto h^{2.6}$ scaling from the amplitude of the lensing potential power spectrum [see Eq.~(\ref{eq:lensAmp})]. Its shape at least partially explains the shift in this parameter plane when the lensing is included. 

\subsection{The impact of priors on $n_s$ and $A_s$ in constraining $h$ in \LCDM}
\label{sec:LCDM_priors}

Given the correlation between $h$ and $A_s$/$n_s$, it is of interest to consider how placing priors on the primordial power spectrum affects $h$. Ref.~\cite{Philcox:2022sgj} imposed $n_s = 0.96 \pm 0.02$ or an 8\% prior on $A_s$ centered on the \textit{Planck} value, $\ln 10^{10} A_s = 3.044 \pm 0.08$. Since here we consider both \LCDM\ and beyond-\LCDM\ models which prefer larger values of $n_s$ when fit to CMB data \cite{Smith:2019ihp,Ye:2021nej,Aloni:2021eaq}, we use the same prior on $A_s$ but a slightly wider prior for $n_s = 0.96 \pm 0.03$. We find good agreement with Ref.~\cite{Philcox:2022sgj} when imposing the same priors, and the specific choice of priors do not affect our overall conclusions.

When imposing  both $A_s$ and $n_s$ priors, we find that the resulting posterior on $h$ decreases from $h=0.687_{-0.05}^{+0.03}$ to $h=0.649\pm0.022$. This significant downward shift is not surprising, given  that $h$ is anti-correlated with both $A_s$ and $n_s$ as discussed above, and that these priors are larger than the values preferred by `All' of the data (see Fig.~\ref{fig:LCDM_comp} and \ref{fig:LCDM_ns_vs_h}). When imposing these priors, we find that the value of $h$ is in $3.3\sigma$ tension with \shoes. The tension level is slightly stronger than that reported in Ref.~\cite{Philcox:2022sgj} because we impose both priors at the same time. 

\subsection{The role of $\Omega_m$ in constraining $h$}

The scaling equations discussed in Sec.~\ref{sec:scalings} and the right-hand contours in Fig.~\ref{fig:h_degen}, indicate that the shape of both the FS data and the CMB lensing potential power spectrum constrain various combinations of $\Omega_m h^p$, where $1 \lesssim p \lesssim 2$. These constraints, along with a prior on $\Omega_m$, provides a constraint on $h$. 

The red/blue dashed curves in the top right panel shows the mean and $\pm 1 \sigma$ of $\omega_m = \Omega_m h^2$ and $\Omega_m h$, respectively. The rough agreement indicates that some combination of $\Omega_m h^p$, with $1 \lesssim p \lesssim 2$, plays a role in constraining $h$. Since the $\Omega_m h^p$ constraint comes from several aspects of the measurements with slightly different dependencies-- baryonic effects ($\Omega_m h^2$), the logarithmic enhancement of the $k>k_{\rm eq}$ part of the galaxy power spectrum ($\Omega_m h$), the peak of the lensing potential power spectrum ($\Omega_m h^{1.33}$)-- we expect the degeneracy between $h$ and $\Omega_m$ to be less well-defined. The bottom right panel shows that, as with FS+BBN, some combination of $\Omega_m h^p$, with $1 \lesssim p \lesssim 2$, continues to play a role in constraining $h$ in the FS+BBN+CMBLens analysis.

These scaling equations indicate that if the prior on $\Omega_m$ decreases then the inferred value of $h$ will increase. 
So far we have used a prior on $\Omega_m$ from PanPlus: $\Omega_m = 0.338 \pm 0.018$ \cite{Brout:2022vxf}. This value is $\sim 2 \sigma$ larger than the value of $\Omega_m$ inferred from the uncalibrated BAO and CMB  measurements of the projected sound horizon: $\Omega_m = 0.3 \pm 0.01$ \cite{Lin:2021sfs}. 

Replacing the PanPlus prior on $\Omega_m$ with the BAO/CMB angular prior allows us to explore how information about $\Omega_m$ impacts the $r_{s,d}$-marginalized \LCDM\ posterior on $h$. As expected with the lower $\Omega_m$ the mean value of $h$ increases from $h=0.687^{+0.030}_{-0.050}$ to $h=0.734^{+0.033}_{-0.063}$. This demonstrates that at least part of the apparent tension in $\Lambda$CDM with SH0ES comes from the relatively high value of $\Omega_m$ favored by Pantheon+. If we also impose the $A_s$/$n_s$ prior then the posterior distribution for $h$ increases from $h= 0.649\pm 0.022$ to $h=0.688^{+0.018}_{-0.021}$. With the $A_s/n_s$ prior we can see that the change in the $\Omega_m$ prior leads to a $1.3 \sigma$ shift in the mean of $h$.

\section{Constraints in beyond-\LCDM\ models}
\label{sec:beyondLCDM}
% \subsection{Preliminary study: Highlighting the correlation with $r_{s,d}$}
We first establish that without marginalizing over $r_{s,d}$ the three beyond-\LCDM\ models that we consider have the expected effect on the value of $r_{s,d}$. 
\begin{figure}[h!] 
   \centering
   \includegraphics[width=0.8\columnwidth]{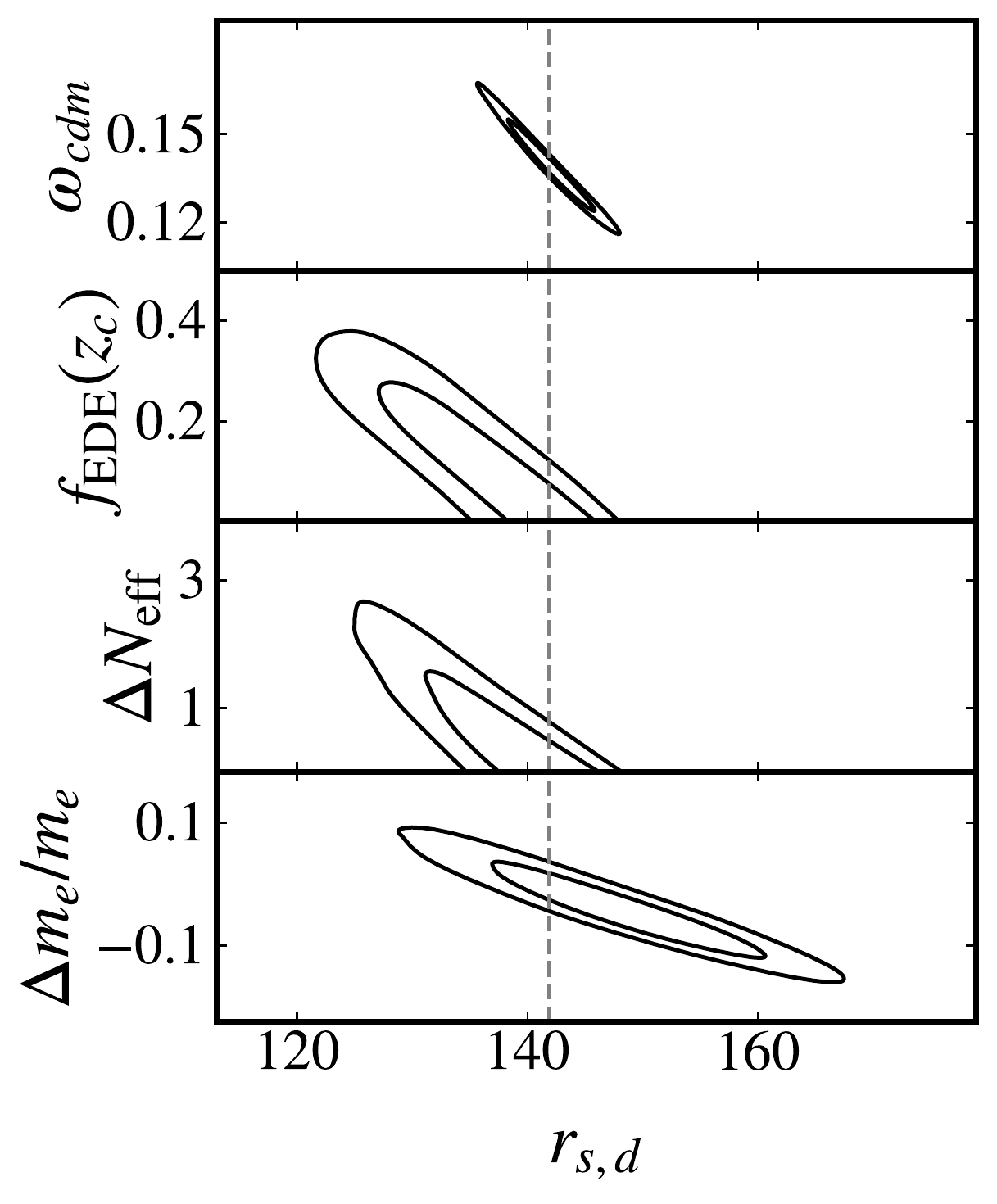} 
   \caption{The 2D posterior distribution for \LCDM\ and the three beyond-\LCDM\ models we consider using `All' of the data (without marginalizing over $r_{s,d}$). The dashed gray line is the mean value for \LCDM. We can see that as the beyond-\LCDM\ model parameters vary the inferred value of $r_{s,d}$ varies, as expected.}
   \label{fig:vs_rsd}
\end{figure}

Fig.~\ref{fig:vs_rsd} shows that the three beyond-\LCDM\ models affect $r_{s,d}$ as expected. In particular, $f_{\rm EDE}(z_c)$-- which controls the maximum contribution that the EDE field makes to the total energy density-- is only able to \emph{increase} the pre-recombination value of $H$, and therefore it can only lead to a \emph{decrease} in $r_{s,d}$. Variations in the number of massless neutrinos, $\Delta N_{\rm eff}>0$, can only cause a decrease in $r_{s,d}$ from its \LCDM\ value (shown by the vertical dashed line). The Thomson scattering cross-section scales as $1/m_e^2$, so a larger electron mass leads to a decrease in the scattering rate, which in turn causes the baryons to decouple earlier than they would have. Therefore as $m_e$ increases, $z_{d}$ increases, leading to a decrease in $r_{s,d}$ (see Eq.~\ref{eq:rs}). 

\begin{figure}[h!] 
   \centering
   \includegraphics[width=0.8\columnwidth]{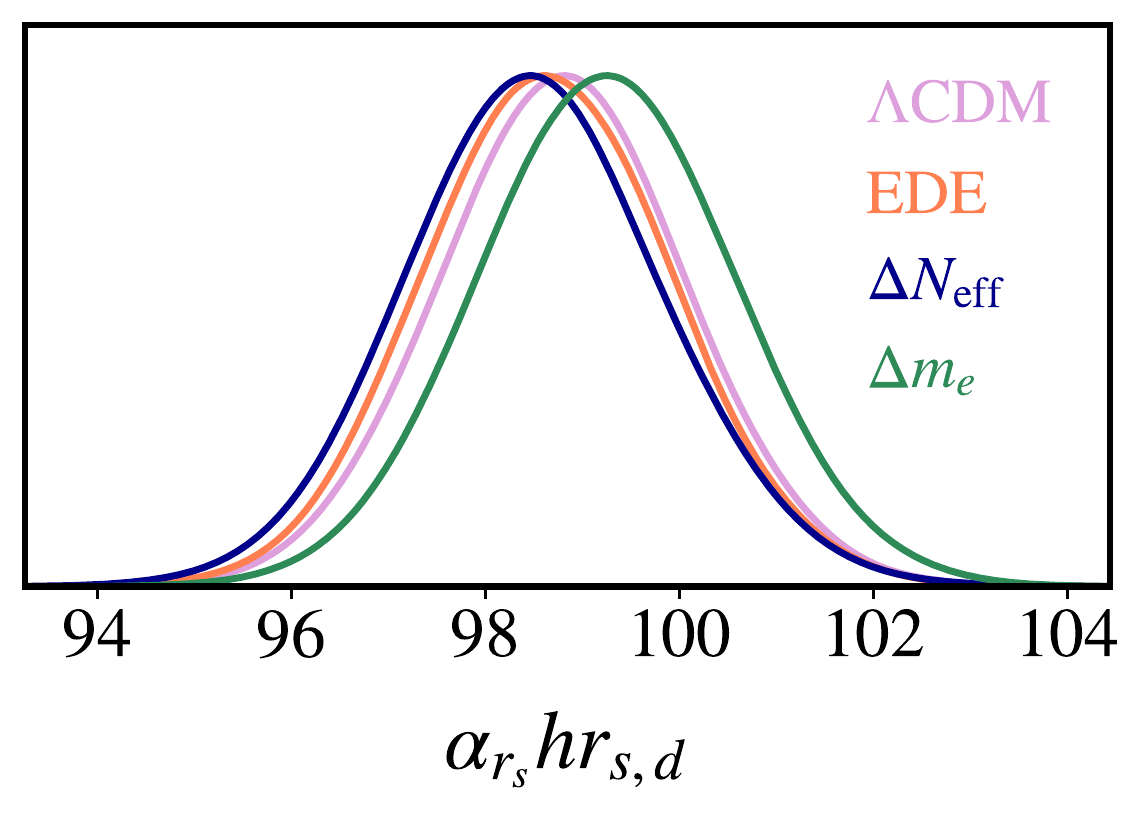} 
   \caption{When marginalizing over $r_{s,d}$ we are effectively measuring the product $\alpha_{r_s} h r_{s,d}$. Here we can see that all four cosmological models produce statistically identical posterior distributions for this product when using `All' of the data.}
   \label{fig:prod}
\end{figure}

\begin{table*}[!t]
\def\arraystretch{1.0}
    \scalebox{1.0}{
    \begin{tabular}{|l|c|c|c|c|} 
        \hline
        \hline
         & \LCDM & EDE & $\Delta N_{\rm eff}$ & $\Delta m_e$ \\
        \hline
        w/o $r_{s,d}$ marg & $0.697^{+0.014}_{-0.016}$  & $0.736^{+0.027}_{-0.036}$ & $0.724^{+0.021}_{-0.030}$ & $0.671^{+0.031}_{-0.040}$ \\
        \hline
        with $r_{s,d}$ marg & $0.687^{+0.030}_{-0.050}$ & $0.708^{+0.038}_{-0.049}$ & $0.699^{+0.034}_{-0.050}$ & $0.684^{+0.031}_{-0.049}$  \\
        \hline
        PanPlus$\rightarrow \theta_{s,d}^{\rm BAO/CMB}$ & $0.734^{+0.033}_{-0.063}$ & $0.748^{+0.038}_{-0.046}$ & $0.739^{+0.035}_{-0.052}$  & $0.716^{+0.032}_{-0.038}$ \\
        \hline
        +$A_s$ \& $n_s$ prior & $0.649\pm 0.022$ & $0.687^{+0.030}_{-0.039}$ & $0.681^{+0.027}_{-0.038}$ & $0.647^{+0.019}_{-0.023}$ \\
        \hline
        PanPlus$\rightarrow \theta_{s,d}^{\rm BAO/CMB}$ & $0.688^{+0.018}_{-0.021}$ & $0.737^{+0.032}_{-0.039}$ & $0.726^{+0.028}_{-0.037}$  & $0.688\pm 0.019$ \\
        \hline
        \hline
    \end{tabular} }
    \caption{The mean and $\pm 1\sigma$ uncertainties of $h$ in the four models we explore. The `PanPlus' prior is $\Omega_m = 0.338 \pm 0.018$ and the uncalibrated BAO and CMB  measurements of the projected sound horizon, `$\theta_{s,d}^{\rm BAO/CMB}$', prior is $\Omega_m = 0.3 \pm 0.01$. When we replace the $\Omega_m$ prior we apply it to the analysis described in the above row.}
    \label{tab:h_beyondLCDM}
\end{table*}

\subsection{$r_{s}$-marginalized constraints on $H_0$ beyond $\Lambda$CDM}

We provide the marginalized constraints on $h$ for \LCDM\ and the three beyond-\LCDM\ models we consider in Table \ref{tab:h_beyondLCDM}. 

The oscillation frequency of the BAO is equal to $r_{s,d}$. Since we marginalize over the product $\alpha_{r_s} k$ within the BAO, and observations are in angular/redshift space, the directly measured quantity is $\alpha_{r_s} h r_{s,d}$, and therefore should be relatively stable between the different models we have analyzed. In Fig.~\ref{fig:prod} we show that for the three cosmological models we analyze, this combination is relatively unchanged, as expected. This provides evidence that our marginalization over $\alpha_{r_s}$ is correct even in these extended cosmologies. This is discussed further in Appendix \ref{app:check}. 

\begin{figure}[h!] 
  \centering
  \includegraphics[width=\columnwidth]{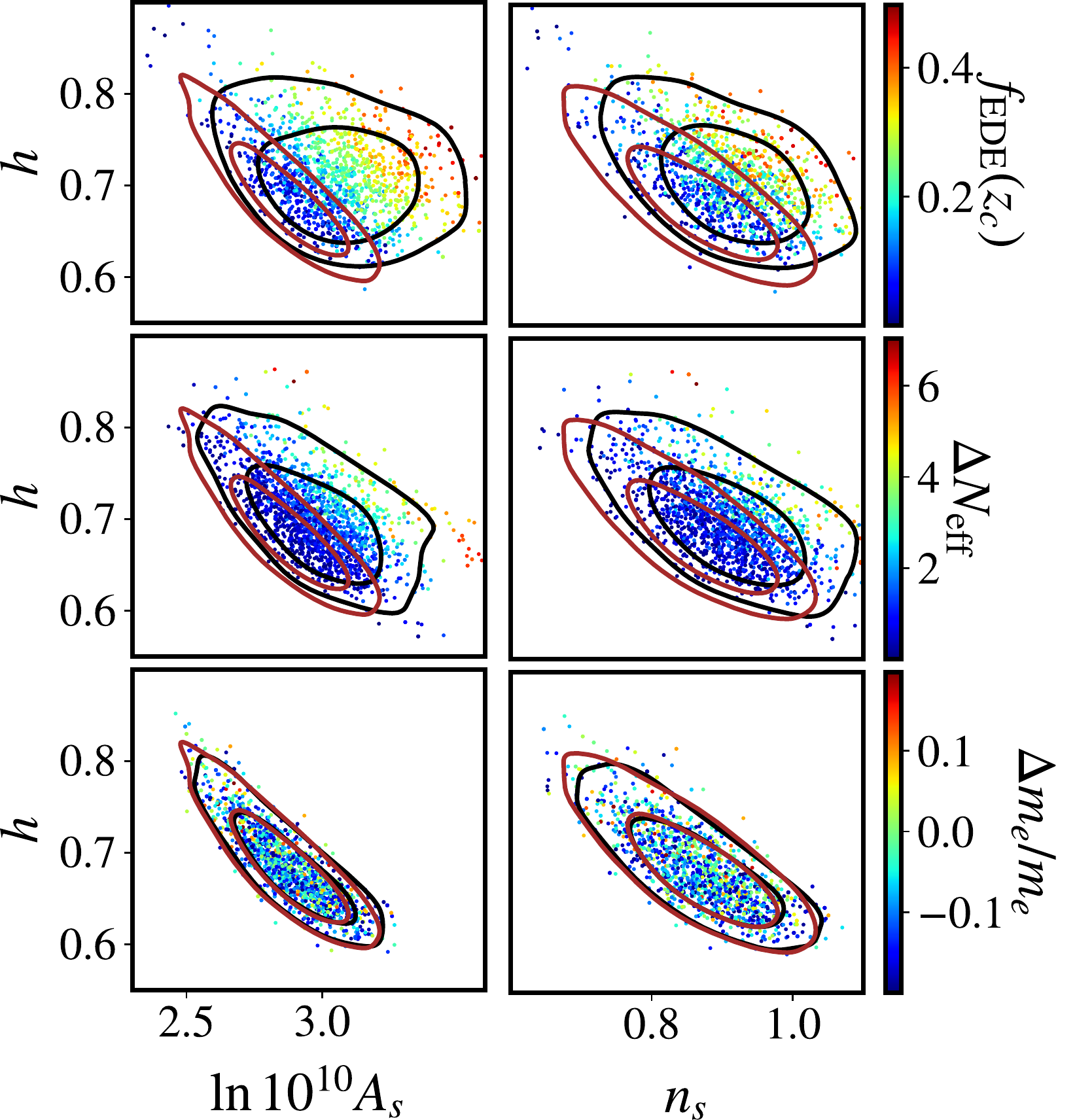} 
  \caption{The $h$ vs.~$\ln 10^{10} A_s$ and $h$ vs.~$n_s$ 2D posterior distributions for the three beyond-\LCDM\ models we consider from the $r_{s}$-marginalized analyses. The brown contours show the \LCDM\ constraints. In EDE and $\Delta N_{\rm eff}$ we can see that additional parameter space is opened which allows for a larger value of $h$ at larger values of $A_s$ with a corresponding increase in the model parameter (shown in the color bars). On the other hand, when varying $m_e$, we obtain contours statistically identical to \LCDM. }
  \label{fig:vs_As}
\end{figure}

The main result of this section is shown in Fig.~\ref{fig:vs_As}. 
There we can see how both $\Delta N_{\rm eff}$ and EDE produce similar posteriors in the $h$ vs. $\ln 10^{10} A_s/n_s$ plane, whereas the varying $m_e$ model is qualitatively different, and similar to what we obtain in \LCDM\ (shown in the brown contour in the bottom plot). The color bars of Fig.~\ref{fig:vs_As} show that the larger values of $\Delta N_{\rm eff}$ and $f_{\rm EDE}(z_c)$ open up a new degeneracy, allowing for a simultaneous increase in $h$, $A_s$, and $n_s$. 
\begin{figure}[h!] 
   \centering
   \includegraphics[width=0.8\columnwidth]{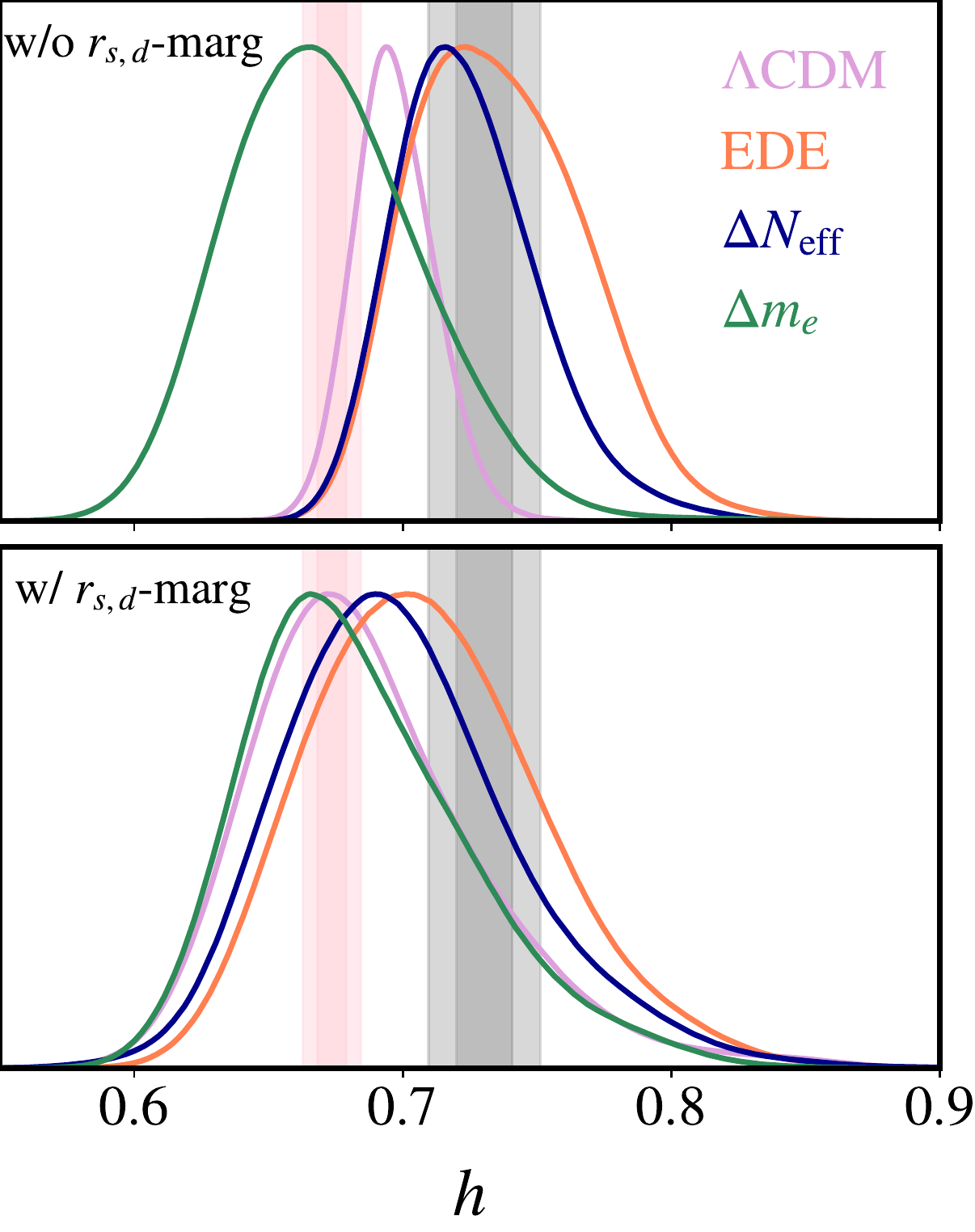}
   \caption{The 1D posterior distribution for $h$ without (top) and with (bottom) marginalizing over $r_{s,d}$. The top panel shows that the inferred value of $h$, which is dominated by information about the sound horizon, varies between the four cosmological models. When marginalizing over $r_{s,d}$ the posteriors become similar, with EDE shifted to a slightly larger value of $h$. The gray bands indicate the \shoes~ value of $h=0.73 \pm 0.01$ and the pink bands the \textit{Planck} value of $h=0.6736 \pm 0.0054$.}
   \label{fig:h_alone}
\end{figure}
This is due to the fact that unlike $\Delta m_e$, both EDE and $\Delta N_{\rm eff}$ introduce additional energy density with significant pressure support. This leads to a suppression of the growth of matter perturbations, leading to a degeneracy with the primordial power spectrum-- i.e., $A_s$ and $n_s$-- for these models, allowing these parameters to take on larger values than they do in \LCDM\ and $\Delta m_e$. 

\begin{figure}[h!] 
   \centering
   \includegraphics[width=\columnwidth]{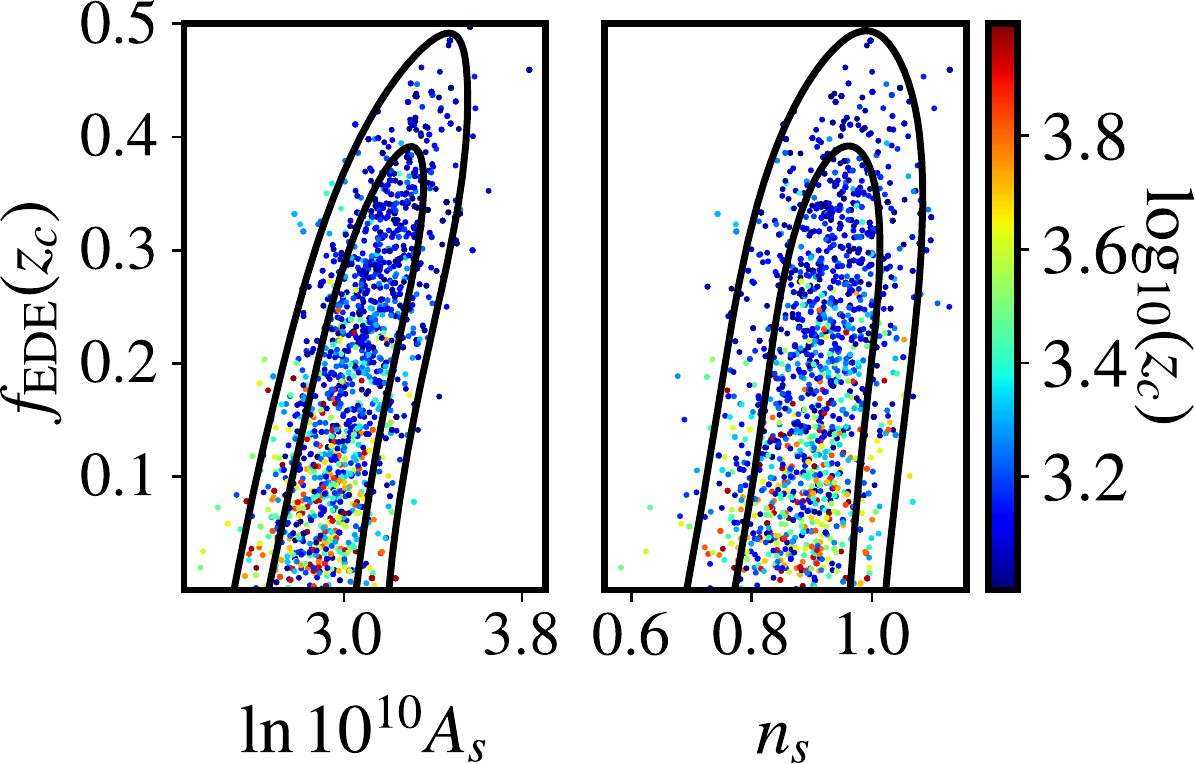}
   \caption{The 2D posterior distribution for $A_s$ and $n_s$ vs.~$f_{\rm EDE}(z_c)$ from the $r_{s}$-marginalized analysis. We can see that the largest values of $n_s$ and $A_s$ in EDE occur when $z_c$ is at the lower end of its prior range. This additional parameter allows EDE to achieve a 1D posterior distribution for $h$ that is slightly larger than in $\Delta N_{\rm eff}$.}
   \label{fig:EDE_zc}
\end{figure}

One can see how this increase in parameter space affects the 1D marginalized posterior distribution for $h$ in Fig.~\ref{fig:h_alone}. Without marginalizing over $r_{s,d}$ (top panel) the posterior distribution for $h$ varies significantly between the different models. When marginalizing over $r_{s,d}$ both EDE and $\Delta N_{\rm eff}$ are shifted to larger values of $h$ than \LCDM\ and $\Delta m_e$. It is also clear that EDE opens up more parameter space volume than $\Delta N_{\rm eff}$. An important distinction between the physics of these two models is that the additional neutrino energy density has an effect throughout radiation domination whereas the additional energy density in EDE is only briefly relevant. This leads to a different scale dependence of their effects and different degeneracies with $A_s$ and $n_s$, which allows EDE to achieve a larger posterior for $A_s$ and $n_s$, as shown in Fig.~\ref{fig:EDE_zc}, with larger values of $A_s/n_s$ corresponding to smaller values of $\log_{10} z_c$. 

We note that even if the EDE/$\Delta N_{\rm eff}$ posteriors for $h$ were not shifted to larger values, the width of all of the $h$ posteriors can easily account for the \textit{Planck} and \shoes-inferred values. As such, none of these analysis rule out these models as resolutions to the Hubble tension. 

\subsection{Impact of $n_s$ and $A_s$ priors }

Just as in \LCDM, given the negative degeneracy between $A_s/n_s$ and $h$ in an $r_{s}$-free analysis (see Fig.~\ref{fig:vs_As}) any priors on these parameters will lead to a significant change in the 1D posterior distribution for $h$. 
\begin{figure}[t!] 
   \centering
   \includegraphics[width=0.8\columnwidth]{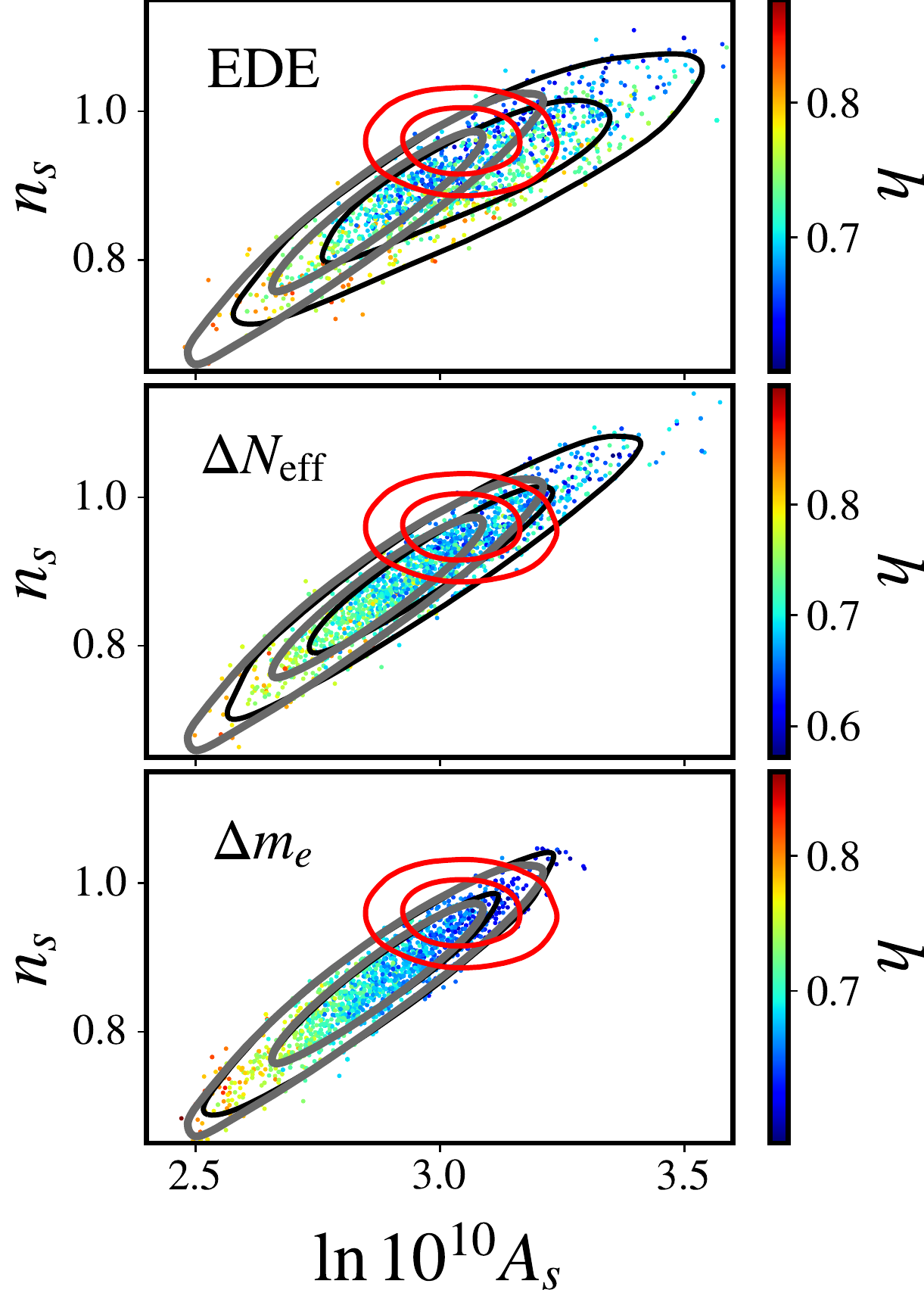} 
   \caption{The 3D posterior distribution for `All' of the data (from the $r_{s}$-marginalized analysis) in the $n_s$ vs.~$\ln 10^{10} A_s$ plane along with color coded points indicating the corresponding value of $h$ for the three beyond-\LCDM\ models we explore. The gray contours show the results in \LCDM, the red contours show the CMB priors we place on $A_s$ and $n_s$.}
   \label{fig:As_vs_ns}
\end{figure}
The result of including a CMB prior on $A_s$ and $n_s$ is shown in the second from the bottom row of Table \ref{tab:h_beyondLCDM}. There we can see that the degeneracies introduced by EDE and $\Delta N_{\rm eff}$ lead to a $\sim 1\sigma$ shift in $h$ to higher values compared to \LCDM\ and $\Delta m_e$. 

We can better understand how the $A_s$ and $n_s$ prior affects these analysis by examining Fig.~\ref{fig:As_vs_ns}. There we show the 3D posterior for $n_s$, $\ln 10^{10} A_s$, and $h$. The gray contour shows the $n_s$ vs.~$\ln 10^{10} A_s$ posterior distribution in \LCDM~ and the red contour shows the CMB prior on $A_s$ and $n_s$. The EDE and $\Delta N_{\rm eff}$ panels clearly show that these models open new parameter space to allow for larger values of $A_s$ and $n_s$ at correspondingly larger values of $h$. When placing a prior on $A_s$ and $n_s$ this additional volume leads to a 1D posterior distribution for $h$ which is shifted to larger values (i.e., cyan and yellow points) than in $\Delta m_e$ or \LCDM. 

\subsection{Impact of the $\Omega_m$ priors }

Replacing the PanPlus prior on $\Omega_m = 0.338 \pm 0.018$ with $\theta_{s,d}^{\rm BAO/CMB}$, $\Omega_m = 0.3 \pm 0.01$, results in an increase in the 1D posterior for $h$ for all models with or without the $A_s/n_s$ prior, as shown in Table \ref{tab:h_beyondLCDM}. 

\begin{figure}[h!] 
  \centering
  \includegraphics[width=\columnwidth]{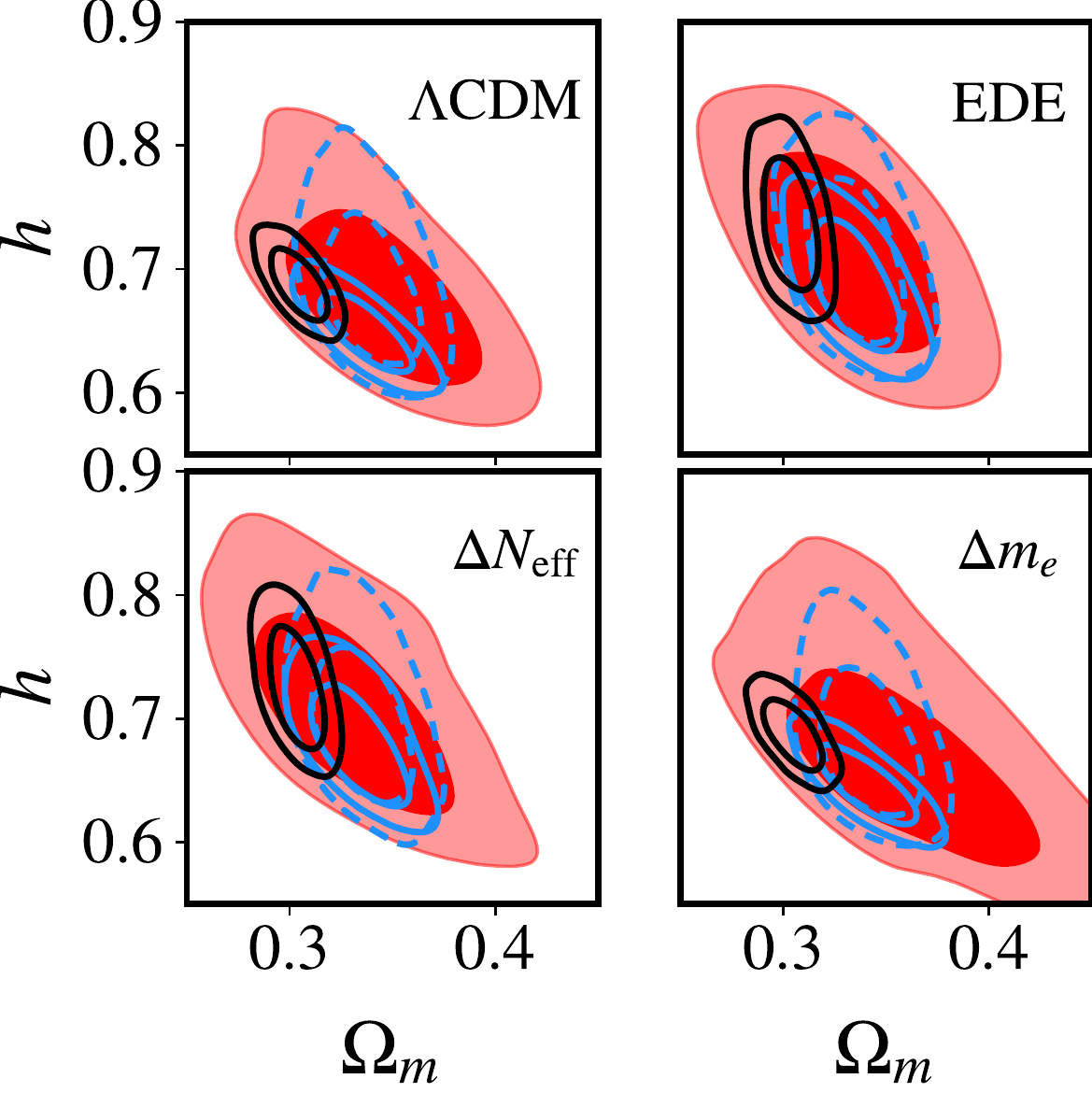} 
  \caption{The $h$ vs.~$\Omega_m$ degeneracy $r_{s}$-marginalized analyses. The filled red contours show constraints using FS+BBN+CMBLens (i.e., without a prior on $\Omega_m$). The open dashed blue contours show the constraints when we include the PanPlus prior on $\Omega_m$ and the open solid blue contours add the effects of the CMB-inpsired priors on $A_s$ and $n_s$. The black contours show the constraints when we keep the priors on $A_s$ and $n_s$ but replace the PanPlus prior with the one from the uncalibrated BAO and CMB  measurements of the projected sound horizon. }
  \label{fig:Om_prior}
\end{figure}

The red contours in Fig.~\ref{fig:Om_prior} show the $h$ vs.~$\Omega_m$ degeneracy in all four models using FS+BBN+CMBLens (i.e., no prior on $\Omega_m$). One can see that a negative degeneracy between $h$ and $\Omega_ m$ is present in all four models we consider. The dashed blue contours show the posterior when we include the PanPlus prior, the solid blue contours further include the CMB-inspired priors on $A_s/n_s$, and the black contours show the posteriors when PanPlus is replaced with $\theta_{s,d}^{\rm BAO/CMB}$. One can see that, when the prior on $\Omega_m$ decreases, the contours shift along the $h$/$\Omega_m$ degeneracy leading to larger values of $h$ (see the blue vs. black contours in Fig.~\ref{fig:Om_prior}). In addition, this figure clearly shows how the inclusion of the $A_s/n_s$ prior significantly reduces the range of $h$ for both \LCDM\ and $\Delta m_e$, but has a much smaller effect for EDE and $\Delta N_{\rm eff}$ (see the dashed blue vs. solid blue contours in Fig.~\ref{fig:Om_prior}).

Note that in our analysis we fixed the sum of the masses of the neutrinos to their minimum value (0.06 eV). We expect that also allowing the neutrino mass to vary, as done in Ref.~\cite{Philcox:2022sgj}, would make the constraints on $h$ even weaker.

\section{Conclusions}
\label{sec:conclusions}

Full-shape information from measurements of galaxy clustering are poised to contribute important cosmological information when investigating beyond \LCDM\ models. Therefore it is important to clarify what aspects of these measurements are driving the constraints. 

The constraining power on $h$ predominately comes from the BAO sensitivity to the sound horizon, and the same is true of measurements of the CMB. This has lead to the development of a number of beyond-\LCDM\ models which change the value of the sound horizon in order to address the Hubble tension. In order to further test these models, it is of interest to develop new analysis methods that extract information about $h$ from observations which are based on pre-recombination physics without relying on the value of the sound horizon. 

The full-shape analysis of measured galaxy power spectra can provide such a data set \cite{Philcox:2020xbv}. By marginalizing over the sound horizon and using a BBN prior on $\omega_b$, SNeIa prior on $\Omega_m$, and the measured CMB lensing from \textit{Planck}, the inference of $h$ relies on the amplitude and broad-band shape of the small-scale power spectrum. Previous work has focused on the sensitivity of these data to $k_{\rm eq} = \Omega_m h$, along with a SNeIa-inspired prior on $\Omega_m$, as the main source of sensitivity to $h$. 
Here we have demonstrated that the sensitivity is also driven by the amplitude of the small-scale power spectrum. 
As a result, beyond-\LCDM\ models which are degenerate with $A_s/n_s$ have the ability to affect the $r_{s}$-free value of $h$. 

This also has potential implications for using the extended BAO parameter set presented in Ref.~\cite{Brieden:2021edu} and known as `ShapeFit'. In this approach, the standard BAO and redshift space distortion parameters are augmented with a parameter that measures the slope of the galaxy power spectrum at $k_{\rm slope} = 0.03 h {\rm Mpc}^{-1}$. It has been shown that this extended parameter set is competitive with full-shape analysis of \LCDM\ \cite{Brieden:2022lsd}. Since we show here that constraints to the amplitude of the galaxy power spectrum play an important role when considering beyond-\LCDM\ models, it will be interesting to check whether ShapeFit will be able to capture some of the important effects of these models.

\begin{figure}[t!] 
  \centering
  \includegraphics[width=\columnwidth]{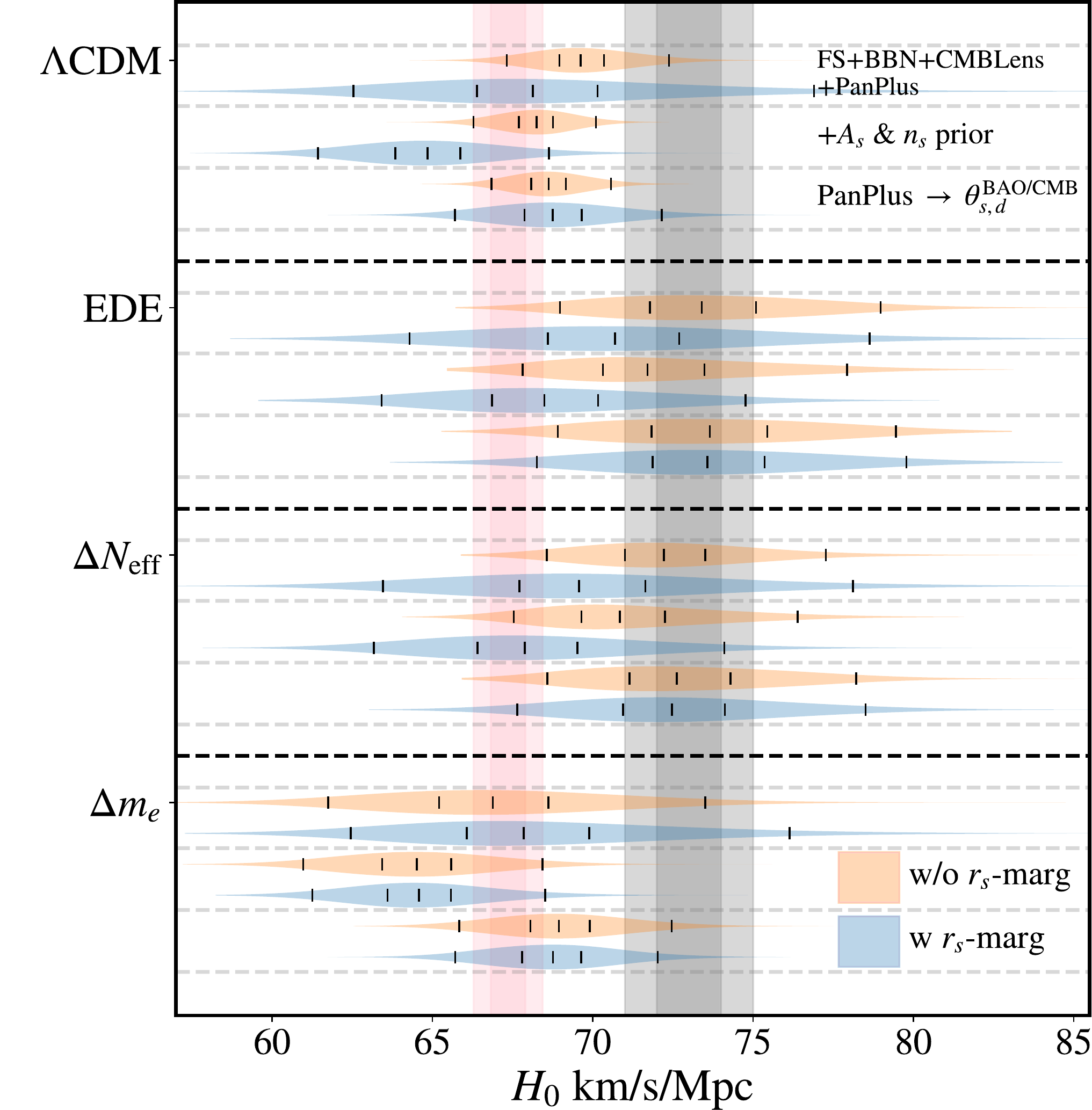}
  \caption{A full summary of our results. The orange/blue posteriors show the results with/without marginalizing over $r_{s,d}$. For each cosmological model we show analyses with three different choices of priors. We also show the \textit{Planck} constraint to $H_0$ as the vertical pink band and the \shoes\ constraint in the gray band.}
  \label{fig:whisker2}
\end{figure}

We find that beyond-\LCDM\ models which introduce additional energy density with significant pressure support lead to increased values of $h$ in an $r_s$-independent analysis. 
This is due to the ways in which these models suppress the growth of structure and are therefore degenerate with the amplitude of the clustering. 
Since the amplitude of the small-scale galaxy power spectrum and lensing potential power spectrum play a central role in determining the $r_{s}$-free value of $h$, models which attempt to address both the Hubble and $S_8$ tensions through a suppression of small-scale power \cite{Allali:2021azp,Clark:2021hlo,Joseph:2022jsf} may be particularly interesting to consider in light of the analysis presented here. 

We have also explored how various priors on cosmological parameters affect these conclusions. 
When using a CMB-inspired prior on $A_s$ and $n_s$ we found that the model-dependence of these results are even more stark, with EDE and $\Delta N_{\rm eff}$ giving posteriors for $h$ which are $\sim 1\sigma$ larger than in \LCDM. 
However, the $\Delta m_e$ model, which only affects recombination, has a posterior for $h$ that is statistically identical to the result in \LCDM.  
Additionally, we have emphasized the role played by the Pantheon+ prior on $\Omega_m$ in driving the low-$h$ constraints. 
Replacing the Pantheon+ prior on $\Omega_m = 0.338 \pm 0.018$ with one from the uncalibrated BAO and CMB  measurements of the projected sound horizon, $\Omega_m = 0.30 \pm 0.01$, leads to a shift to higher values of $h$ for all models, with EDE and $\Delta N_{\rm eff}$ still $\sim 1 \sigma$ larger than \LCDM.  
The posteriors for $h$ are listed in Tab.~\ref{tab:h_beyondLCDM}.

We conclude that the Hubble constant inferred from these data depends on both the model and the choice of priors on the cosmological parameters. 

Our analysis also allows us to determine whether a comparison between the $H_0$ posteriors with and without marginalizing over $r_{s,d}$ in \LCDM\ provides a robust internal consistency test for physics beyond \LCDM. 
A summary of these results is shown in Fig.~\ref{fig:whisker2}. Using FS+BBN+CMBLens+PanPlus, one can see that without any prior on $A_s$ and $n_s$, the agreement in \LCDM\ is better than $1\sigma$. The agreement is slightly worse ($\sim 2-2.5\sigma$) once the $A_s$ and $n_s$ priors are included, with a shift in the means of $\Delta H_0\sim 3$km/s/Mpc.\footnote{Using the same priors on $n_s$ and the sum of the neutrino masses as in Ref.~\cite{Philcox:2022sgj} we find a similar result, where without (with) marginalizing over $r_{s,d}$, $h=0.682_{-0.012}^{+0.011}$ ($h=0.652_{-0.026}^{+0.022}$)} Keeping the $A_s/n_s$ prior and changing the $\Omega_m$ prior to the uncalibrated BAO and CMB  measurements of the projected sound horizon brings the $H_0$-values back into excellent agreement. 

Given these results, at a minimum we conclude that the consistency of $H_0$ with and without $r_{s,d}$-marginalization in \LCDM\ depends on the choice of priors on the cosmological parameters. In addition, when the \LCDM\ posteriors are consistent, we do not find any indication that the beyond-\LCDM\ models are in tension with the data. 
Given this, our results indicate that with current data the internal consistency test proposed in Refs.~\cite{Farren:2021grl,Philcox:2022sgj} is inconclusive.

The results presented here complement those that are presented in Ref.~\cite{Simon:2022adh}. There we show that the BOSS full-shape analysis using both \texttt{PyBird} and \texttt{CLASS-PT} do not rule out the EDE resolution to the Hubble tension. In light of Ref.~\cite{Simon:2022lde}, it will be useful to perform an analysis similar to what we have done here but using \texttt{PyBird}, since this code relies on a different choice of EFT priors and BOSS power-spectrum measurements. Indeed,  the constraints from these two codes may differ up to $\sim 1 \sigma$ for \LCDM\ due (mostly) to the impact of priors \cite{Simon:2022lde}.
However, we do not expect the overall conclusions to change, as we have identified physical effects at play in driving degeneracies between $h$ and other parameters. 

Current galaxy clustering measurements are not precise enough to rule out or favor beyond \LCDM\ models which address the Hubble tension. However, unlike CMB lensing \cite{Baxter:2020qlr}, there are several near-future galaxy surveys which will significantly improve constraints on $h$ independent of the sound horizon upon BOSS DR12 (e.g., DESI \cite{DESI:2016fyo}, Euclid \cite{EUCLID:2011zbd}, VRO \cite{2009arXiv0912.0201L}). 
The work presented here highlights the ways in which beyond-\LCDM\ models which address the Hubble tension may affect the value of $h$ even in an $r_{s}$-free analyses. 
 
\begin{acknowledgements}

We thank Pierre Zhang for contributions at early stages of this work, and his comments and insights throughout the project, Adam Riess, Jose Bernal and Blake Sherwin for helpful comments on the draft, and Eric Jensen and Gerrit Farren for useful conversations. We thank Antony Lewis for help with \texttt{getdist}, Oliver Philcox for help with \texttt{CLASS-PT}, and Adam Riess for providing us with the Pantheon+ likelihood. This work used the Strelka Computing Cluster, which is run by Swarthmore College. TLS is supported by NSF Grant No.~2009377, NASA Grant No.~80NSSC18K0728, and the Research Corporation. This project has received support from the European Union’s Horizon 2020 research and innovation program under the Marie Skodowska-Curie grant agreement No 860881-HIDDeN.

\end{acknowledgements}

\appendix

\section{Derivation of the \LCDM\ parameter scaling equations}
 \label{app:scaling}

\subsection{Approximate scalings for the galaxy power spectrum}

It is helpful to recall the basic physics that determines the small-scale ($k>k_{\rm eq}$) form of the matter power spectrum. Roughly speaking, dark matter modes with $k>k_{\rm eq}$ enter the horizon during radiation domination and experience a large Hubble friction, significantly limiting their growth. Once the universe becomes matter dominated all of those modes are able to collapse, growing proportional to $a/a_{\rm eq}$. This scaling gets modified in detail since the dark matter perturbations do grow logarithmically with scale factor during radiation domination \cite{1974A&A....37..225M}, giving an amplitude of the galaxy power spectrum 
\begin{eqnarray}
P_{\rm gal}(k>k_{\rm eq}) &\propto& b^2 R_c^2 g(z)^2A_s (a/a_{\rm eq})^{2} \\&\times& [1+\ln(4 a_{\rm eq}/a_k)]^2 \left(\frac{k}{k_p}\right)^{n_s-1}(h/k)^3,\nonumber\\
&=& b^2 f_b\left[\frac{\omega_b}{\omega_{cdm}}\right]\Omega_m^{0.25} A_s a^2 \Omega^2_m h^4  \label{eq:gal} \\&\times& \left[1+\ln\left(\frac{4k/h}{\Omega_m h}\right)\right]^2 \left(\frac{k}{k_p}\right)^{n_s-1}(h/k)^3,\nonumber
\end{eqnarray}
where $b$ is the linear galaxy bias, $R_c \equiv \omega_{cdm}/\omega_m = 1-\omega_b/\omega_m$ is the baryon suppression \cite{Bernal:2020vbb}, horizon crossing occurs when $k=a_k H(a_k)$, $g(z)^2 \propto \Omega_m^{0.25}$ is the growth function at $z \sim 0.3-0.6$, and $k_p$ is the pivot scale (usually chosen to be $k_p = 0.05\ {\rm Mpc}^{-1}$). We note that information about the bias comes from redshift space distortions and the use of informative priors. The second line shows the explicit dependence on $h$ in \LCDM. During radiation domination we have $a_k = 100\ {\rm km/s/Mpc}/c \sqrt{\omega_r}h/k$. Using the fact that $a_{\rm eq} \equiv \omega_r/\omega_m$ we can write $a_{\rm eq}/a_k \simeq k/k_{\rm eq}$. We can see that for $k>k_{\rm eq}$ the logarithmic term enhances the amplitude. A more careful treatment shows that the logarithmic term is $\ln[k/(8k_{\rm eq})]$, so for $k_{\rm eq} \sim 0.01\ h{\rm Mpc}^{-1}$ and $k_{\rm max} = 0.4\ h{\rm Mpc}^{-1}$ we get an enhancement of power at the smallest scales of a factor of $\sim 7$ \cite{Eisenstein:1997ik,Dodelson:2003ft}. This enhancement gives the sensitivity to $k_{\rm eq}$.

The correlation of the monopole and quadrupole moments of the galaxy clustering power spectrum gives us redshift space distortion information which provides sensitivity to the product of the growth rate, $f(z)$, and the variance of mass fluctuations in spheres of radius $R = 8\ {\rm Mpc} h^{-1}$ ($\sigma_8^2$). First, from Ref.~\cite{Planck:2015mym} we have
\begin{equation}
    \sigma_8^2 \propto A_s (a/a_{\rm eq})^2 \Omega_m^{0.25} (k_{\rm eq} h^{-1})^{-1.4} \omega_m^{0.45},
\end{equation}
where the dependence on $\Omega_m$ comes from the growth function around the BOSS DR12 redshift bins ($z\sim 0.5$).  
In $\Lambda$CDM, the growth rate is approximately \cite{1992ARA&A..30..499C}
\begin{equation}
    f(z\sim0.5) \propto \Omega_m^{0.6}.
\end{equation}

\subsection{Approximate scaling for the lensing potential power spectrum}

Since the \textit{Planck} inferred lensing potential power spectrum provides measurements between $8 \leqslant L \leqslant 400$ \cite{Planck:2018lbu}, there are two relevant quantities in the CMB lensing: position of the peak $\ell^{\phi \phi}_{\rm peak}$ and the amplitude of high $L$ power spectrum, $L^4C_L^{\phi \phi}$.

First, the peak of the spectrum is set by $\theta_{\rm eq}$ at $z\sim 2$ \cite{Lewis:2006fu}, so that $\ell^{\phi \phi}_{\rm peak} \propto \Omega_m^{0.75} h\omega_r^{-0.5}$.

Second, the CMB lensing potential power spectrum also has sensitivity to $k_{\rm eq}$. A rough approximation to the combination of parameters measured by estimates of the lensing potential power spectrum is given by \cite{Planck:2015mym} 
\begin{eqnarray}
    L^4 C_L^{\phi \phi} &\propto& A_s \ell_{\rm eq}^{2} \omega_m^{0.3},\\
    &=& A_s h^{2.6}\Omega_m^{3.5}
\end{eqnarray}
where $\ell_{\rm eq} \equiv \chi_{\rm dec} k_{\rm eq}$, $\chi_{\rm dec}$ is the comoving distance to photon decoupling, the power law index for $\omega_m$ is fit around $L \simeq 200$, and the primordial power spectrum was taken to be scale invariant. The product $A_s \ell_{\rm eq}^2$ can be simply understood: the gravitational potential power spectrum is nearly scale invariant up until $k_{\rm eq}$, at which point it becomes small. The number of collapsed halos of size $r \sim k^{-1}$ that a CMB photon passes by is given by $\chi_*/r \sim k\chi_*$, where $\chi_*$ is the comoving distance to the surface of last scattering, and the typical halo potential is $\sim A_s^{1/2}$. Since $\ell_{\rm eq} = k_{\rm eq} \chi_*$ gives the largest number of halos along the line of sight, the overall amplitude of the deflection power spectrum (which, in turn, is proportional to the lensing potential power spectrum) is proportional to $A_s \ell_{\rm eq}^2$ \cite{2010GReGr..42.2197H}.  Additionally, it is straightforward to show that the angular scale of matter radiation equality at the CMB is $\ell_{\rm eq} \propto \Omega_m^{0.6} h \omega_r^{-0.5}$. 

\section{The effect of removing the galaxy power spectrum peak}
\label{app:peak}

\begin{figure}[h!] 
  \centering
  \includegraphics[width=0.8\columnwidth]{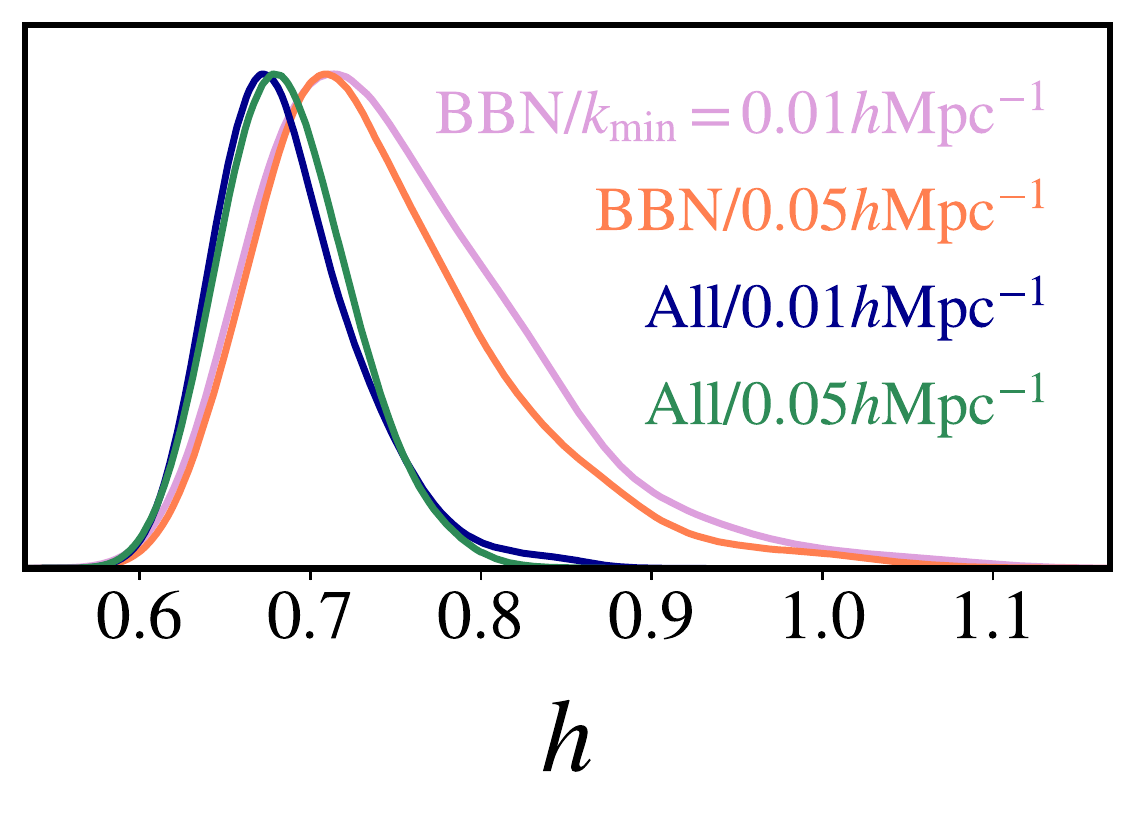} 
  \caption{1D posterior distribution for $h$ in \LCDM\ when marginalizing over $r_{s,d}$ with two different choices for the minimum wavenumber used in the galaxy power spectra multipoles. When using $k_{\rm min} = 0.05\ h{\rm Mpc}^{-1}$ we have removed all information about the location of the peak of the galaxy power spectrum. The statistical equivalence for these two values of $k_{\rm min}$ demonstrates that the location of the peak of the galaxy power spectrum does not play a significant role in constraining $h$.}
  \label{fig:h_kmin}
\end{figure}

To demonstrate that the location of the peak of the galaxy power spectrum is not playing a role in constraining $h$, we performed an analysis with $k_{\rm min} = 0.05\ h {\rm Mpc}^{-1}$ for the galaxy power spectrum multipoles. This choice is $\sim 5$ times larger than $k_{\rm eq}$, fully removing the peak from the data. The resulting 1D posterior for $h$ is shown in Fig.~\ref{fig:h_kmin}. We can see that the posterior is statistically identical to our fiducial choice of $k_{\rm min} = 0.01\ h {\rm Mpc}^{-1}$. This is not surprising given the fact that the fiducial $k_{\rm min}$ is just slightly less than $k_{\rm eq}$. We note that the signal to noise in the lowest measured modes is smallest since at the largest scales we have the fewest independent measurements. 

We note that, although galaxy power spectra may not be able to probe scales large enough to measure the peak, future HI surveys will have enough coverage \cite{Cunnington:2022ryj}. 

\section{Checking the BAO smoothing algorithm}
\label{app:check}

\begin{figure}[h!] 
   \centering
   \includegraphics[width=\columnwidth]{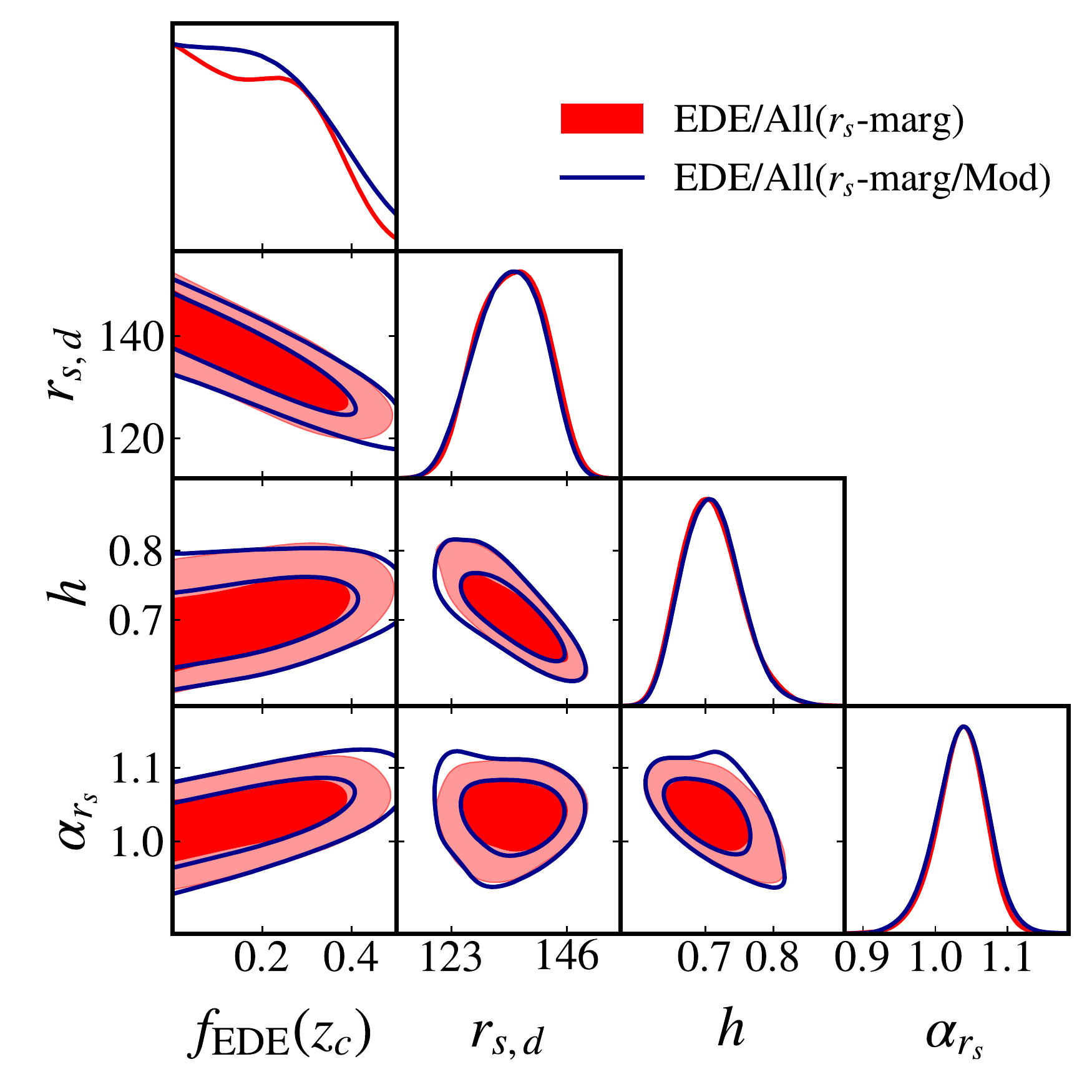}
   \caption{A triangle plot comparing the posterior distribution between the two smoothing algorithms.}
   \label{fig:EDE_comp}
\end{figure}

Fig.~\ref{fig:vs_rsd} indicates that part of the parameter space may not be modeled correctly. As discussed in Ref.~\cite{Chudaykin:2020aoj} the BAO smoothing algorithm used in \texttt{CLASS-PT} is constructed to work well for $130\ {\rm Mpc} \leqslant r_{s,d} \leqslant 170\ {\rm Mpc}$. Clearly the \LCDM\ MCMCs have samples which are slightly beyond the lower end of this range. The algorithm performs a sine-transform of the matter power spectrum and, excises the BAO bump, interpolates between the two smooth regions on either side, and then inverse transforms back to Fourier space. The excision of the BAO bump is done using fixed boundaries, and so will fail if the BAO bump gets close to either of those boundaries. In order to investigate whether this causes an issue at the lower boundary, we modified the algorithm slightly by allowing the boundary to move as the value of $r_{s,d}$ changes. 

Our modified algorithm keeps the distance between the excised points and shifts it linearly with the value of $r_{s,d}$ with the standard value at  $r_{s,d}=150$ Mpc. The original algorithm fixes the region of the real-space correlation function that is excised in order to remove the BAO bump. In terms of the indices they remove all points between $N_{\rm left} = 120$ and $N_{\rm right} = 240$ \cite{Chudaykin:2020aoj}. We have modified the range of indices which are removed so that it is translated as the value of $r_{s,d}$ changes:
\begin{equation}
    N_{\rm left} = 120-20(1-r_{s,d}/150)/(1-120/150),
\end{equation}
and $N_{\rm right} = N_{\rm left} + 120$. We have verified that this algorithm properly excises the BAO bump when $r_{s,d}$ is varied between $110\ {\rm Mpc} \leqslant r_{s,d} \leqslant 170\ {\rm Mpc}$.

The comparison between the standard and modified algorithm for EDE is shown in Fig.~\ref{fig:EDE_comp}. We focus on EDE here since the value of $r_{s,d}$ has the largest range in this model. There we can see that when `All' of the data is included the two methods are nearly identical. We have checked the other cosmological models we consider show similar insensitivity to the change in the broadband/BAO split. 

\section{Verifying the Pantheon+ prior on $\Omega_m$}
\label{app:PanPlus_check}

\begin{figure}[h!] 
   \centering
   \includegraphics[width=\columnwidth]{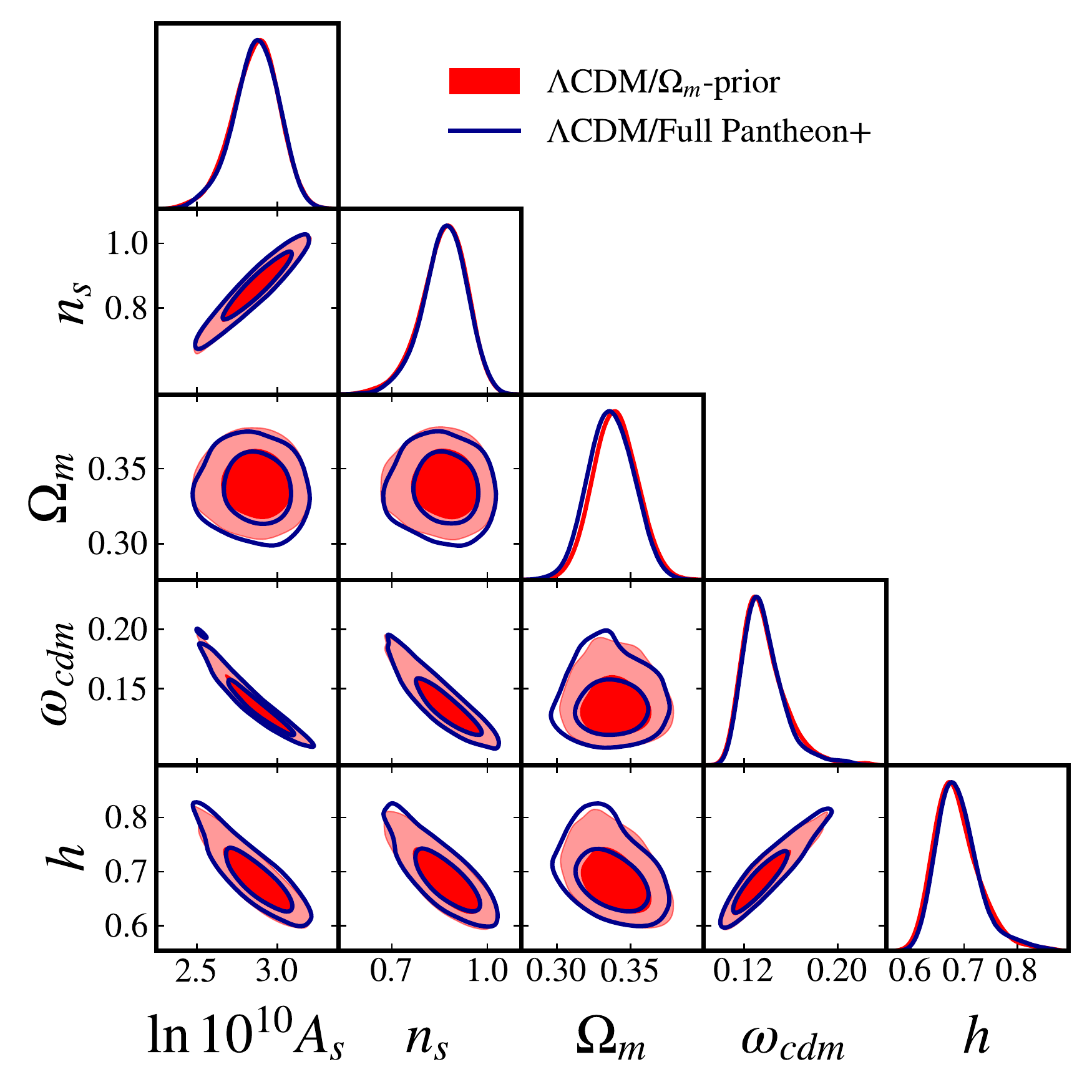}
   \caption{A triangle plot comparing the posterior distribution between the full Pantheon+ likelihood and a prior on $\Omega_m = 0.338 \pm 0.018$.}
   \label{fig:PanPlus_comp}
\end{figure}

Instead of using the full Pantheon+ likelihood we have used a prior on $\Omega_m = 0.338 \pm 0.018$. In order to verify that this prior properly captures all of the aspects of this likelihood we have compared the constraints on \LCDM\, marginalizing over $r_{s,d}$, using the FS+BBN+CMBLens+PanPlus, implementing the full Pantheon+ likelihood vs.~using the prior on $\Omega_m$.

The comparison between these two analyses is shown in Fig.~\ref{fig:PanPlus_comp}. There we can see that the posteriors are nearly identical, verifying our use of the Pantheon+ prior on $\Omega_m$. The conclusions of this comparison also hold in the beyond-\LCDM\ models we consider since they all introduce new physics at or before recombination, and therefore are identical to \LCDM\ in the late universe when SNeIa measurements are made. 

\newpage
\bibliography{biblio}

\end{document}